\documentclass[preprint,authoryear,12pt]{elsarticle}

\usepackage{graphics}
\usepackage{graphicx}
\usepackage{epsfig}

\usepackage{amssymb}
\usepackage{amsthm}

\usepackage{lineno}
\usepackage{color}

\begin{document}

\begin{frontmatter}



\title{
Finite Scale Lyapunov Analysis of Temperature Fluctuations in Homogeneous Isotropic  Turbulence
}


\author{Nicola de Divitiis}

\address{"La Sapienza" University, Dipartimento di Ingegneria Meccanica e Aerospaziale, Via Eudossiana, 18, 00184 Rome, Italy}

\begin{abstract}
This study analyzes the temperature fluctuations in incompressible homogeneous isotropic turbulence through the finite scale Lyapunov analysis of the relative motion between two fluid particles.
The analysis provides an explanation of the mechanism of the thermal energy cascade, leads to the closure of the Corrsin equation, and describes the statistics of the longitudinal temperature derivative through the Lyapunov theory of the local deformation and the thermal energy equation. 
The results here obtained show that, in the case of self-similarity, the temperature spectrum exhibits the scaling laws $\kappa^n$, with
$n \approx{-5/3}$, ${-1}$ and ${-17/3} \div -11/3$ depending upon the flow regime. These results are in agreement with the theoretical arguments of Obukhov--Corrsin and Batchelor and with the numerical simulations and experiments known from the literature. 
The PDF of the longitudinal temperature derivative is found to be a non--gaussian distribution function with null skewness, whose intermittency rises with the Taylor scale P\'eclet number.
This study applies also to any passive scalar which exhibits diffusivity.
\end{abstract}


\end{frontmatter}

\newcommand{\no}{\noindent}
\newcommand{\be}{\begin{equation}}
\newcommand{\ee}{\end{equation}}
\newcommand{\bea}{\begin{eqnarray}}
\newcommand{\eea}{\end{eqnarray}}
\newcommand{\bc}{\begin{center}}
\newcommand{\ec}{\end{center}}

\newcommand{\calr}{{\cal R}}
\newcommand{\calv}{{\cal V}}

\newcommand{\bff}{\mbox{\boldmath $f$}}
\newcommand{\bfg}{\mbox{\boldmath $g$}}
\newcommand{\bfh}{\mbox{\boldmath $h$}}
\newcommand{\bfi}{\mbox{\boldmath $i$}}
\newcommand{\bfm}{\mbox{\boldmath $m$}}
\newcommand{\bfp}{\mbox{\boldmath $p$}}
\newcommand{\bfr}{\mbox{\boldmath $r$}}
\newcommand{\bfu}{\mbox{\boldmath $u$}}
\newcommand{\bfv}{\mbox{\boldmath $v$}}
\newcommand{\bfx}{\mbox{\boldmath $x$}}
\newcommand{\bfy}{\mbox{\boldmath $y$}}
\newcommand{\bfw}{\mbox{\boldmath $w$}}
\newcommand{\bfk}{\mbox{\boldmath $\kappa$}}

\newcommand{\bfA}{\mbox{\boldmath $A$}}
\newcommand{\bfD}{\mbox{\boldmath $D$}}
\newcommand{\bfI}{\mbox{\boldmath $I$}}
\newcommand{\bfL}{\mbox{\boldmath $L$}}
\newcommand{\bfM}{\mbox{\boldmath $M$}}
\newcommand{\bfS}{\mbox{\boldmath $S$}}
\newcommand{\bfT}{\mbox{\boldmath $T$}}
\newcommand{\bfU}{\mbox{\boldmath $U$}}
\newcommand{\bfX}{\mbox{\boldmath $X$}}
\newcommand{\bfY}{\mbox{\boldmath $Y$}}
\newcommand{\bfK}{\mbox{\boldmath $K$}}

\newcommand{\bfrho}{\mbox{\boldmath $\rho$}}
\newcommand{\bfchi}{\mbox{\boldmath $\chi$}}
\newcommand{\bfphi}{\mbox{\boldmath $\phi$}}
\newcommand{\bfPhi}{\mbox{\boldmath $\Phi$}}
\newcommand{\bflambda}{\mbox{\boldmath $\lambda$}}
\newcommand{\bfxi}{\mbox{\boldmath $\xi$}}
\newcommand{\bfLambda}{\mbox{\boldmath $\Lambda$}}
\newcommand{\bfPsi}{\mbox{\boldmath $\Psi$}}
\newcommand{\bfomega}{\mbox{\boldmath $\omega$}}
\newcommand{\bfOmega}{\mbox{\boldmath $\Omega$}}
\newcommand{\bfeps}{\mbox{\boldmath $\varepsilon$}}
\newcommand{\bfepsn}{\mbox{\boldmath $\epsilon$}}
\newcommand{\bfzeta}{\mbox{\boldmath $\zeta$}}
\newcommand{\bfkappa}{\mbox{\boldmath $\kappa$}}
\newcommand{\itPsi}{\mbox{\it $\Psi$}}
\newcommand{\itPhi}{\mbox{\it $\Phi$}}
\newcommand{\bint}{\mbox{ \int{a}{b}} }
\newcommand{\ds}{\displaystyle}
\newcommand{\Sum}{\Large \sum}


\section{\bf Introduction}
 \label{intro}

This work adopts the finite--scale Lyapunov theory for studying the temperature fluctuations in incompressible homogeneous isotropic turbulence in an infinite fluid domain. The study is mainly motivated by the fact that, in isotropic turbulence, the temperature spectrum $\Theta(\kappa)$ exhibits several scaling laws $\kappa^n$ in the different wavelength ranges depending on $R$ and $Pr$ (\cite{Corrsin_2, Obukhov, Batchelor_2, Batchelor_3}), where $R$ and $Pr$ are Taylor scale Reynolds number and Prandtl number, respectively.
This is due to the peculiar connection between temperature fluctuations, fluid deformation and velocity field, whose effect varies following $R$ and $Pr$. 

\begin{figure}[h]
	\centering
 \hspace{0.mm}        \includegraphics[width=0.60\textwidth]{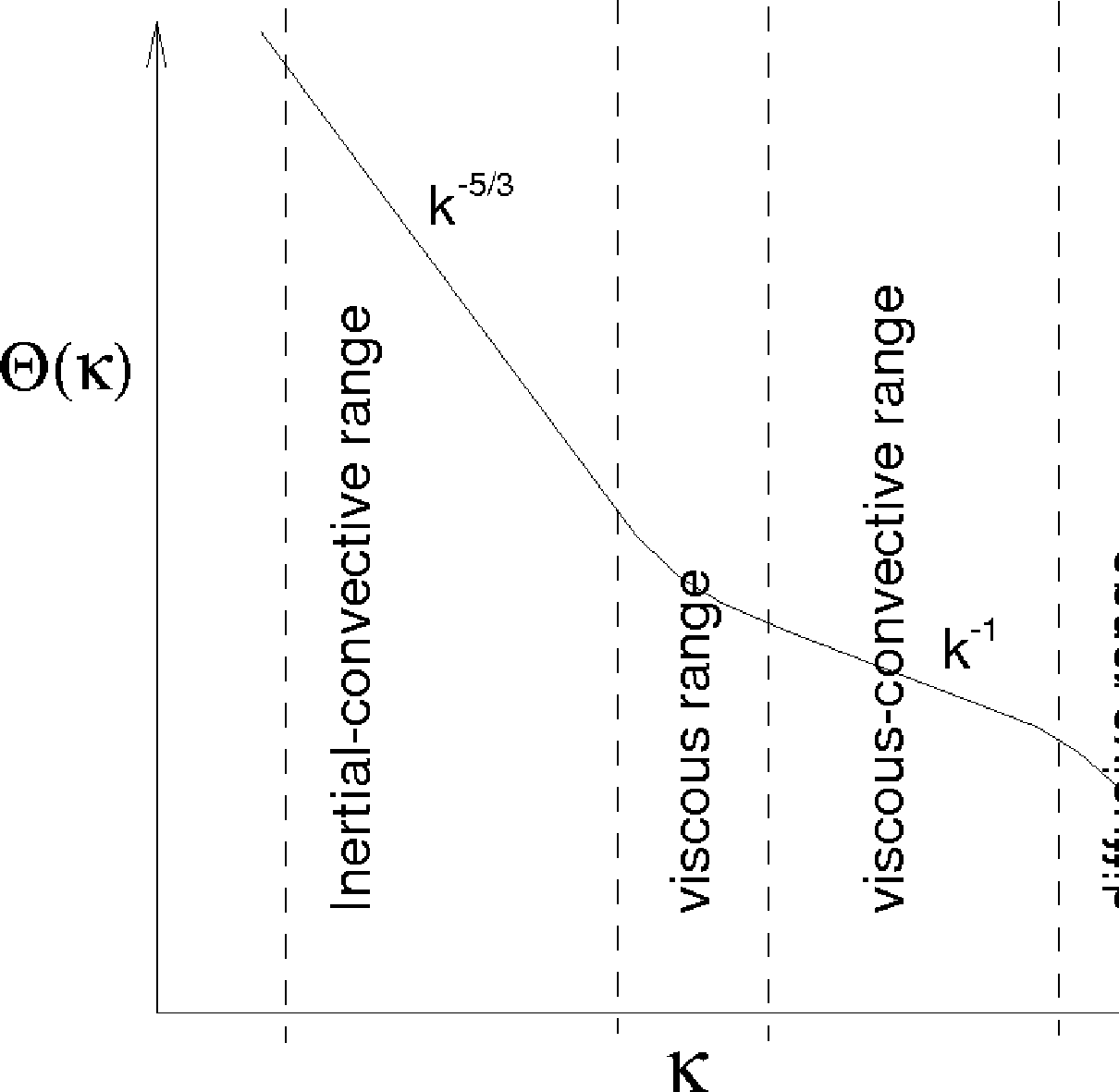}
\caption{Scheme of the subranges of the temperature spectrum at high Prandtl numbers}
\label{figura_0}
\end{figure}

For large values of $R$ and $Pr$, \cite{Corrsin_2} and \cite{Obukhov} argumented, through the dimensional analysis, that $\Theta(\kappa) \approx \kappa^{-5/3}$ in the so--called inertial--convective subrange (see Fig. \ref{figura_0}).
\cite{Batchelor_2} considered the isotropic turbulence at high Prandtl number, when $R$ is assigned. There, the author assumed that, at distances less than the Kolmogorov scale, the temperature fluctuations are mainly related to the strain rate associated to the smallest scales of the velocity field. As the result, he showed that $\Theta \approx \kappa^{-1}$ in the so--called viscous--convective interval, a region where the
scales are less than the Kolmogorov length (see Fig. \ref{figura_0}). 
Different experiments dealing with the grid turbulence (\cite{Gibson,  Mydlarski}) and calculations of the temperature spectrum through numerical simulations (see \cite{Donzis} and references therein) confirm that $\Theta(\kappa)$ follows such these scaling laws.

On the contrary, when $Pr$ is very small, the high fluid conductivity determines quite different situations with respect to the previous ones.
\cite{Batchelor_3} analyzed the small--scale variations of temperature  fluctuations in the case of large conductivity, and found that $\Theta(\kappa) \approx \kappa^{-17/3}$, whereas \cite{Rogallo} calculated  the temperature spectra through numerical simulations of a passive scalar convected by a velocity field with zero correlation time. \cite{Rogallo} showed that, when the kinetic energy spectrum follows the Kolmogorov law $E(\kappa) \approx \kappa^{-5/3}$, the temperature spectrum varies according to $\Theta(\kappa) \approx \kappa^n$, with $n \approx - 11/3$.

According to the experiments of grid turbulence, temperature and velocity correlations are linked with each other when $Pr =O(1)$, whereas the decay rate and characteristic scales depend on the initial conditions.
Specifically, \cite{Mills} obtained very important data about the air turbulence behind a heated grid. They carried out several measurements of nearly isotropic fluctuations of velocity and temperature at different distances from the grid, and recognized that
$f_\theta \simeq f$ and  $p_*\simeq k$, where $f_\theta$ and $f$ are 
temperature and velocity correlations respectively,  $p_*$ is the triple correlation temperature--velocity, and $k$ is the longitudinal triple velocity correlation.
Later, \cite{Warhaft} experimentally showed that spectrum shape and decay rate depend upon the initial conditions and that the mechanical--thermal time scale ratio tends to a value close to the unity.

Other important characteristics of $\Theta(\kappa)$ is the self--similarity. This is related to the idea that the combined effect of thermal and kinetic energy cascade in conjunction with conductivity and viscosity, 
makes the temperature correlation similar in the time.
This property was theoretically studied by George (see \cite{George1, George2} and references therein) which showed that the decaying isotropic turbulence reaches the self--similarity, where $\Theta(\kappa)$ is scaled by the Taylor microscale whose current value depends on the initial condition.
Recently, \cite{Antonia} studied the temperature structure functions in decaying homogeneous isotropic turbulence and found that the standard deviation of the temperature, as well as the turbulent kinetic energy, follows approximately the similarity over a wide interval of length scales. There, the authors used this approximate similarity to calculate the third--order correlations and found satisfactory agreement between measured and calculated functions.

{\color{black} Very important advances, regarding other properties of passive scalars in fully developed turbulence, were recently made (\cite{Fereday, Schekochihin, Doering, Tran07, Tran08,  Burton}).

\cite{Fereday} studied the decay in a large--scale flow and
discussed the relation between the decay obtained by the Lagrangian stretching theories
and that calculated with the numerical simulations.
Among the other things, the authors determined that the PDF of a passive scalar exhibits algebraic tails, with an exponent of about -3 in a given interval of dimensionless scalar concentration, with a cutoff due to the fluid diffusivity.
For what concerns the decay models of a passive scalar, \cite{Schekochihin} analyzed the case with single--scale random velocity field, and showed that, if there exists separation between flow scale and the box size, the decay rate is the result of the turbulent diffusion of the box--scale.
Later, \cite{Doering} studied the mixing efficiency of a passive scalar subject to a steady and inhomogeneous source, advected by a statistically homogeneous and isotropic incompressible velocity field. The authors found that the mixing efficiency is limited by the values of $Pr \ R$ and by specific characteristics of the source,
and that the scaling laws of the bounds at high $Pr \ R$ depend on the length scales of the source.
\cite{Tran07}, in an article dealing with the scalar diffusion in shear flows, determined an upper bound for the decay rate of the temperature standard deviation in the case of shear flows with bounded velocity gradients, where the initial temperature distribution is supposed to be a smooth function of the space co--ordinates. Thereafter, \cite{Tran08} analyzed the evolution of the temperature gradient and showed, thanks to the hypothesis of finite velocity gradient, that the square of temperature gradient and its decay rate are both bounded. 
Next, \cite{Burton} extended the nonlinear large--eddy simulation method to conditions with moderate and very high Schmidt numbers, and, among the other things, provided the instantaneous field of scalar--energy at viscous--convective scales at high Schmidt--numbers.}

\bigskip

From a theoretical point of view, the properties of $\Theta(\kappa)$ can be investigated through
its evolution equation. 
$\Theta(\kappa)$ is the Fourier--Transform of $f_\theta$ which varies
according to the Corrsin equation (\cite{Corrsin_1}).
This latter includes $G$, a term  responsible for the thermal energy cascade that is directly related to the triple correlation $p_*$, thus the Corrsin equation is not closed.
As $G$ depends also on the velocity fluctuations,  the Corrsin equation requires 
the knowledge of $f$, thus it must be solved together to the von K\'arm\'an--Howarth equation. On this argument, some work has been written.
For instance, \cite{Baev} (and references therein) studied temperature and kinetic energy spectra adopting a closure model based on the gradient hypothesis, which incorporates empirical constants.
Nevertheless, to the author's knowledge, the estimation of $\Theta(\kappa)$ based on the 
theoretical analysis of the closure of von K\'arm\'an–-Howarth and Corrsin equations 
has not received due attention.

This is the motivation of the present work, whose main objective is to propose the closure of the Corrsin equation and a description of the statistics of the temperature derivative.
The present study is based on the finite--scale Lyapunov theory used by  
\cite{deDivitiis_1, deDivitiis_2}.
Here, this Lyapunov theory gives $G$ in function of $f$ and of $\partial f_\theta / \partial r$, and describes also the statistics of the temperature gradient through the analysis of the local strain and the canonical decomposition of temperature and velocity in terms of proper stochastic variables. The mathematical expression of $G$ is obtained considering that it is frame invariant, thus $G$ can be calculated in a
convenient frame of reference. The adoption of the finite--scale Lyapunov basis is revealed to be a convenient choice for determining $G$.
For what concerns the von K\'arm\'an--Howarth equation, the analytical closure proposed by  \cite{deDivitiis_1} is here used.
Through the self--similarity,
the closed von K\'arm\'an--Howarth and Corrsin equations are reduced to be an ordinary differential system. This latter is numerically solved for several values of $R$ and $Pr$, and the obtained results show that the temperature spectrum exhibits scaling laws depending on $R$ and $Pr$, in agreement with the experimental and theoretical data of the literature 
(\cite{Corrsin_2, Obukhov, Rogallo, Mills, Gibson}).
As far as the statistics of the longitudinal temperature gradient is concerned, it is represented by non--gaussian PDF with null skewness and a Kurtosis greater than three whose value rises with the P\'eclet number $Pe = Pr \ R$.

\bigskip

\section{\bf Background: Corrsin equation}

For the sake of convenience, the procedure to obtain the Corrsin equation is here renewed.

Velocity and temperature are considered to be statistically homogeneous isotropic fields, 
and viscosity $\nu$ and thermal conductivity $k$ are assigned quantities.
The equations of the temperature fluctuation $\vartheta$ in two points
${\bf x} \equiv (x, y, z)$ and ${\bf x}' \equiv {\bf x} + \bf r$, are
\bea
\begin{array}{l@{\hspace{-0.cm}}l}
\ds \frac{\partial \vartheta}{\partial t} 
+ \frac{\partial \vartheta}{\partial x_k} u_k 
- \chi \frac{\partial^2 \vartheta}{\partial x_k \partial x_k} =0
\end{array}
\label{P}
\eea
\bea
\begin{array}{l@{\hspace{-0.cm}}l}
\ds \frac{\partial \vartheta'}{\partial t} 
+ \frac{\partial \vartheta'}{\partial x_k'} u_k' 
- \chi \frac{\partial^2 \vartheta'}{\partial x_k' \partial x_k'} =0
\end{array}
\label{P'}
\eea
being ${\bf r}= (r_x, r_y, r_z)$ the separation vector, 
$\chi = k/(\rho C_p)$ is the fluid thermal diffusivity, and 
$C_p$ is the specific heat at constant pressure. 
As well known, the evolution equation of the temperature correlation is determined 
multiplying Eq. (\ref{P}) and (\ref{P'}) by $\vartheta'$
and $\vartheta$, respectively, and summing the equations (\cite{Corrsin_1}).
The so obtained equation, averaged with respect to the ensemble of
the temperature fluctuations, leads to 
\bea
\ds \theta^2 \frac{\partial  f_\theta}{\partial t} + f_\theta  \frac{d \theta^2}{d t} 
-G
-2 \chi \theta^2 \left( \frac{\partial^2 f_\theta}{\partial r^2}  
+ \frac{2}{r} \frac{\partial f_\theta}{\partial r} \right)  = 0
\label{f_theta}
\eea
where  
\bea
f_\theta = \frac{\langle \vartheta \vartheta'\rangle}{\theta^2}
\eea
is the temperature correlation function,
$\theta= \sqrt{\langle \vartheta^2 \rangle}$ is the standard deviation of
the temperature fluctuations, constant in the space, and
\bea
G = - \frac{\partial}{\partial r_k} \langle \vartheta \vartheta' (u_k'-u_k) \rangle
\label{t cascade}
\eea
The first two terms of Eq. (\ref{f_theta}) express the time variations of $f_\theta$ and
 $\theta$, whereas $G$, arising from the convective terms, provides the mechanism of the
thermal energy cascade where $\partial\langle . \rangle/\partial x'_k$=
$-\partial\langle . \rangle/\partial x_k$=$\partial\langle . \rangle/\partial r_k$. 
 The last term of Eq. (\ref{f_theta})  gives the effects of the thermal diffusion.
Because of isotropy, $G(r)$ is an even function of 
$r\equiv \vert {\bf r} \vert$.
According to \cite{Corrsin_1, Corrsin_2}, $G$ is 
\bea
G = 2 \left( \frac{\partial p_*}{\partial r } + 2 \frac{p_*}{r} \right) \equiv
\frac{2}{r^2}  \frac{\partial }{\partial r } \left( p_* r^2 \right) 
\eea
where $p_*(r)$ is the triple correlation between $u_r$ and the temperatures  
in ${\bf x}$ and ${\bf x}'$, i.e.
\bea
 p_*(r) = \frac{\langle u_r \vartheta \vartheta' \rangle}{\theta^2 u}
\eea
and $u_r$ is the longitudinal component of the fluid velocity.
A property of $G(r)$ is that it does not modify $\theta$, hence 
$G(0)=0$ and $G \approx r^2$ near the origin.
Therefore, $p_* \approx r^3$ when $r \rightarrow 0$, and 
Eq. (\ref{f_theta}) gives the evolution equation for $\theta$ putting $r =0$ (\cite{Corrsin_2})
\bea
\ds  \frac{d \theta^2}{d t} =
-12 \chi \frac{\theta^2}{\lambda_\theta^2} 
\label{theta_dot}
\eea
where $\lambda_\theta$ is the scale of the temperature correlation,
or Corrsin microscale, defined as (\cite{Corrsin_1})
\bea
\lambda_\theta = \sqrt{-\frac{2}{f_\theta''(0)}}
\label{lambda_theta}
\eea
in which the superscript apex denotes the differentiation with respect to $r$.
Hence, the evolution equation of $f_\theta$ is
\bea
\ds  \frac{\partial  f_\theta}{\partial t} 
 -12  \frac{\chi}{\lambda_\theta^2} f_\theta 
-G
-2 \chi  \left( \frac{\partial^2 f_\theta}{\partial r^2}  
+ \frac{2}{r} \frac{\partial f_\theta}{\partial r} \right)  = 0
\label{f_theta 2}
\eea
whose boundary conditions are 
\bea
\begin{array}{l@{\hspace{-0.cm}}l}
\ds f_\theta(0) = 1, \\\\
\ds \lim_{r \rightarrow \infty} f_\theta (r) = 0
\end{array}
\label{bcT}
\eea
Note that Eq. (\ref{f_theta 2}) depends also on the velocity fluctuations 
through $G(r)$ whose analytical form is not given at this stage of the analysis.
Therefore, Eq. (\ref{f_theta 2}) is not closed and provides only a link between $f_\theta$ and $G$. Accordingly, the solutions $f_\theta$ of Eq. (\ref{f_theta 2}) will be related to $f$.

\bigskip

\section{\bf Lyapunov Analysis of thermal energy cascade: closure of Corrsin equation}

This section analyses the thermal energy cascade using
the finite--scale Lyapunov theory adopted in \cite{deDivitiis_1}, 
and proposes the analytical expression for $G$.
For this purpose, consider now Eq. (\ref{t cascade}).
As $G$ is frame invariant, for the sake of convenience, it is expressed in the finite--scale Lyapunov basis $E_\lambda$. This basis is associated with the problem of the relative motion between two fluid particles (\cite{deDivitiis_1}) 
\bea
\begin{array}{l@{\hspace{-0.cm}}l}
\ds \frac{d {\bfrho}}{dt} = {\bf u} ({\bfx}+{\bfrho}, t )- {\bf u} ({\bfx}, t ), \\\\
\ds \frac{d {\bfx}}{dt} = {\bf u} ({\bfx}, t)
\end{array}
\label{fslb}
\eea 
where ${\bfrho}$ gives the relative position between the particles, and $\bf u$ varies according to the Navier--Stokes equations. To define  $E_\lambda$, the solutions ${\bfrho}_1$, ${\bfrho}_2$ and ${\bfrho}_3$ of Eq. (\ref{fslb}) are first considered, whose initial conditions ${\bfrho}_1(0)$, ${\bfrho}_2(0)$ and ${\bfrho}_3(0)$ correspond to mutually orthogonal vectors such that 
$\vert {\bfrho}_1(0) \vert = \vert {\bfrho}_2(0)\vert = \vert {\bfrho}_3(0)\vert$. 
The basis $E_\lambda$ is then obtained through the Gram–-Schmidt orthonormalization
process applied to ${\bfrho}_1(t)$, ${\bfrho}_2(t)$ and ${\bfrho}_3(t)$.

In $E_\lambda$, the velocity difference fluctuation reads as (\cite{deDivitiis_1})
\bea
\Delta {\bf u} \equiv {\bf u}' -{\bf u} = \lambda(r) {\bf r} + {\bfomega}_\lambda \times {\bf r}
+ \bfzeta
\label{du}
\eea  
where
$\lambda$ is the maximal finite--scale Lyapunov exponent (associated with the length $r$), 
defined by
$
\lambda(r) \approx 1/T \int_0^T {d {\bf r}}/{dt} \cdot {\bf r}/r^2 \ dt 
$, and calculated in function of $f$ as (\cite{deDivitiis_1})
\bea
\lambda(r) = \frac{u}{r} \sqrt{2 (1-f)}
\label{lamda}
\eea 
${\bfomega}_\lambda$ is the angular velocity of $E_\lambda$ with respect to the inertial frame of reference $\Re$, and ${\bfzeta} \equiv (\zeta_1, \zeta_2, \zeta_3)$, related to the other two exponents,  is expressed in $E_\lambda$ as (\cite{deDivitiis_1})
\bea
\zeta_1 = (\lambda_1 -\lambda) \varrho_1, \ \ \ \
\zeta_2 = (\lambda_2 -\lambda) \varrho_2, \ \ \ \
\zeta_3 = (\lambda_3 -\lambda) \varrho_3
\eea
\begin{figure}[h]
 \hspace{0.mm}        \includegraphics[width=0.60\textwidth]{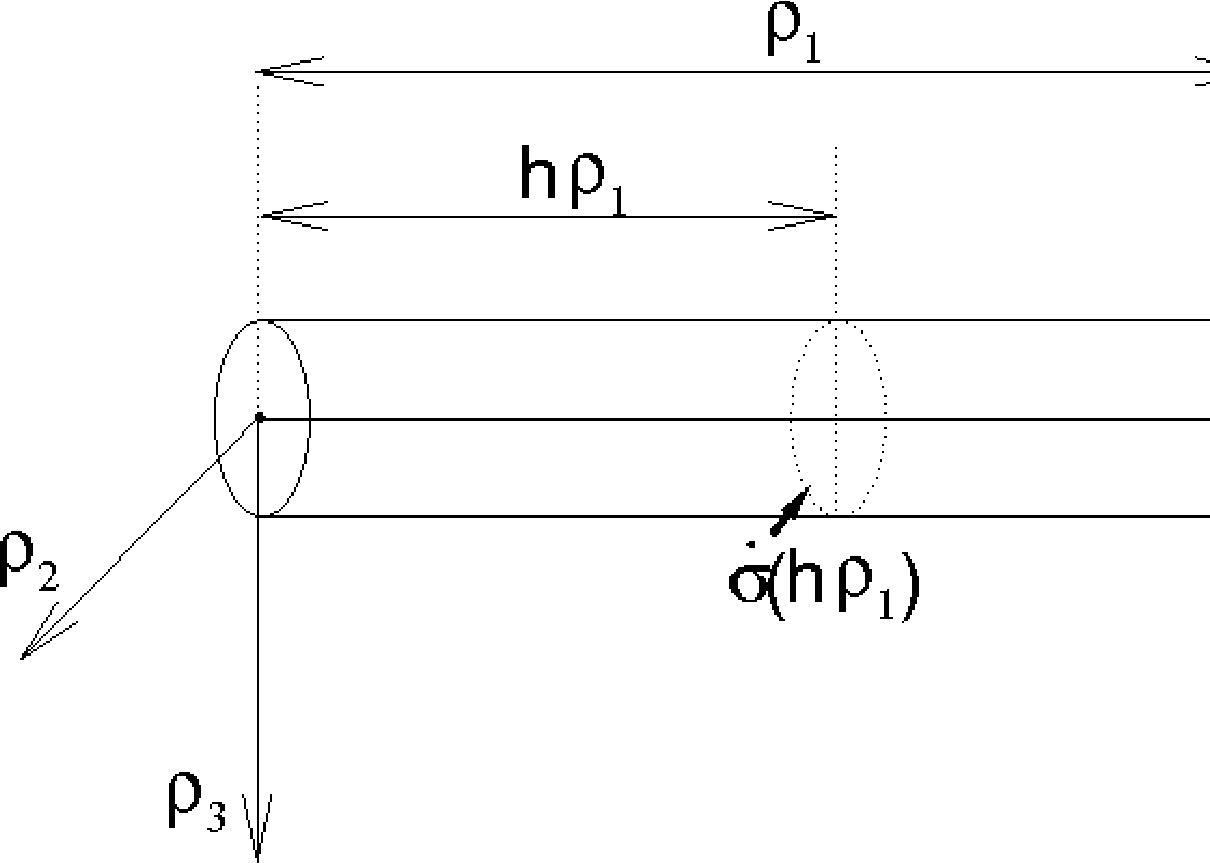}
\caption{Scheme of Finite--scale Lyapunov basis embedded into a material cylinder, 
at a given time}
\label{figura_01}
\end{figure}
where $\lambda_1$, $\lambda_2$ and $\lambda_3$ are the Lyapunov exponents 
associated to the three directions $\varrho_1$, $\varrho_2$ and $\varrho_3$, respectively.
Hence, $\lambda$ is a deterministic quantity, whereas $\bfomega_\lambda$ is
a fluctuating variable related to the relative motion between $E_\lambda$ and $\Re$.
Without lack of generality, the coordinate $\varrho_1$ is supposed to be associated to the maximal exponent, then 
$\lambda_1 \rightarrow \lambda$, $\lambda_2 = \lambda_3 \equiv \lambda_\zeta$, 
$\varrho_1$ diverges being $\vert \varrho_1 \vert >>> \vert \varrho_2 \vert, \vert \varrho_3 \vert$, thus
$\zeta_1 \rightarrow \hat{\zeta_1} \ne 0$ 
\bea
\zeta_1 = \hat{\zeta_1}, \ \ \ \
\zeta_2 = (\lambda_\zeta -\lambda) \varrho_2, \ \ \ \
\zeta_3 = (\lambda_\zeta -\lambda) \varrho_3
\label{zeta canonical}
\eea

The exponents $\lambda_2 = \lambda_3 \equiv \lambda_\zeta$ are determined by the continuity equation.
With reference to Fig. \ref{figura_01}, the equation is written considering, at a given time, the mass balance associated with a material circular cylinder whose axis is parallel to the direction $\varrho_1$
\bea
\ds 2 \frac{\dot{\sigma}}{\sigma} + \frac{ \dot{\varrho}_1}{\varrho_1} = 0
\label{cont1}
\eea
where the dot denotes the differentiation with respect to $t$, 
$\dot{\sigma}$ is evaluated at the coordinate $h \ \varrho_1$, with $h \in (0 ,1)$, 
whereas $\varrho_1$ and $\sigma \equiv \sqrt{\varrho_2^2 + \varrho_3^2}$ are the length and diameter of the cylinder. 
Therefore, $\dot{\varrho}_1/{\varrho}_1$ and $\dot{\sigma}/{\sigma}$ identify $\lambda$ and $\lambda_\zeta$, respectively,  being 
\bea
\ds \lambda_\zeta = - \frac{\lambda}{2}
\eea  
Substituting Eq. (\ref{du}) into Eq. (\ref{t cascade}), and taking into account that $\lambda$ is constant with respect to the operation of statistical average 
(i.e. $\langle \lambda .. \rangle = \lambda \langle .. \rangle$), $G$ is the sum of three addends
\bea
G = G_1 + G_2 + G_3
\label{G r bis}
\eea
where 
\bea
\begin{array}{l@{\hspace{-0.cm}}l}
\ds G_1 = -\frac{\partial}{\partial \varrho_k} \left( \langle \vartheta \vartheta' \rangle \lambda \varrho_k \right), \\\\
\ds G_2 = -\frac{\partial}{\partial \varrho_k} \left( \langle \vartheta \vartheta' \rangle \zeta_k \right), \\\\
\ds G_3 = -\frac{\partial}{\partial \varrho_k} \left( \varepsilon_{k i j} \langle \vartheta \vartheta' \omega_i \rangle  \varrho_j \right) 
\end{array}
\eea
and $\varepsilon_{k i j}$ is the Levi--Civita tensor. 
The expression of $G_1$ is obtained taking into account the isotropy
\bea
G_1 = -\theta^2 \left(  \frac{\partial f_\theta}{\partial r} \lambda r +
 \frac{f_\theta}{r^2} \frac{\partial}{\partial r} \left( r^3 \lambda \right) \right) 
\eea
and $G_2$ is written considering that  
$\lambda(r) {\bf r} +\bfomega_\lambda \times {\bf r} + \bfzeta$ is solenoidal
\bea
G_2= -  \frac{\partial \langle \vartheta  \vartheta' \rangle}{\partial \varrho_k } \zeta_k + 
 \frac{f_\theta \theta^2 }{r^2} \frac{\partial}{\partial r} \left( r^3 \lambda \right)
\eea
where $r^2 = \varrho_1^2+ \varrho_2^2+ \varrho_3^2$.
{\color{black} For what concerns $G_3$, it gives null contribution, as in isotropic turbulence, 
$\langle \vartheta \vartheta' \omega_i \rangle$=$L(\vert {\bfrho} \vert) \varrho_i$, 
where $L(\vert {\bfrho} \vert)$ is an even  function of $\vert {\bfrho} \vert$ (\cite{Batchelor53}).} This implies that $G_3 \equiv 0$,
thus $G$ is
\bea
G = -\theta^2  \frac{\partial f_\theta}{\partial r} \lambda r - 
\frac{\partial \langle \vartheta  \vartheta' \rangle}{\partial \varrho_k} \zeta_k 
\label{G a}
\eea
The first term of Eq. (\ref{G a}) is due to $\lambda >0$ and provides in part the mechanism of the thermal energy cascade, whereas the second one, related to the other two Lyapunov exponents,  
exhibits the opposite sign. To obtain the second term, observe that into 
Eqs. (\ref{zeta canonical}), $\lambda_2 = \lambda_3 = \lambda_\zeta = -\lambda/2$, and this leads to
\bea
\begin{array}{l@{\hspace{-0.cm}}l}
\ds \frac{\partial \langle \vartheta  \vartheta' \rangle}{\partial \varrho_2 } \zeta_2
= \frac{3}{2} \frac{ \partial f_\theta}{\partial r} \lambda \theta^2 \frac{\varrho_2^2}{r}, \ \ \ \ 
\ds \frac{\partial \langle \vartheta  \vartheta' \rangle}{\partial \varrho_3 } \zeta_3
= \frac{3}{2} \frac{ \partial f_\theta}{\partial r} \lambda \theta^2 \frac{\varrho_3^2}{r}
\end{array}
\label{G closure 0}
\eea
Moreover, because of the isotropy ${\partial \langle \vartheta  \vartheta' \rangle}/{\partial \varrho_1 } \hat{\zeta}_1$ must be of the form
\bea
\ds \frac{\partial \langle \vartheta  \vartheta' \rangle}{\partial \varrho_1 } \hat{\zeta}_1
= \frac{3}{2} \frac{\partial f_\theta}{\partial r} \lambda \theta^2 \frac{\varrho_1^2}{r}
\label{G closure 01}
\eea
hence
\bea
\frac{\partial \left\langle \vartheta  \vartheta' \right\rangle}{\partial r_k } \zeta_k 
= -  \frac{3}{2} \theta^2  \frac{\partial f_\theta}{\partial r} \lambda r
\eea
Accordingly, the analytical expression of $G$ is 
\bea
\ds G(r) =  \frac{\theta^2}{2} \frac{\partial f_\theta}{\partial r} \lambda r = 
\theta^2 {u} \sqrt{\frac{1-f}{2}} \  \frac{\partial f_\theta}{\partial r} 
\label{G closure}
\eea
Equation (\ref{G closure}) gives the
proposed closure of the Corrsin equation. 
Its main asset with respect to the other models is that it is not based on the phenomenological assumption, but is derived from a specific finite--scale Lyapunov analysis, under the assumption of incompressible homogeneous isotropic turbulence.
Equation (\ref{G closure}) preserves $\theta$ and describes the mechanism of the
thermal energy cascade. This latter consists of a flow of the thermal energy from
large to small scales whose
effectiveness depends upon $f$ and $f_\theta$.
According to the Lyapunov analysis of \cite{deDivitiis_1}, this mechanism can be viewed in the following manner. 
If, at an initial time $t_0$, a toroidal material volume $\Sigma(t_0)$ is taken, which includes an assigned amount of thermal energy, its geometry and position change according to the fluid motion, and its dimensions will vary to preserve the volume. Choosing $\Sigma$ in such a way that its maximal dimension $R$ increases with $t$, the finite--scale Lyapunov analysis leads to 
$R \approx R(t_0 ) e^{\lambda (t -t_0)}$.  
That is, the thermal energy, initially enclosed into $\Sigma(t_0)$, at the end of the fluctuation is contained into $\Sigma(t)$ whose dimensions are changed with respect to the initial time $t_0$.
Therefore, the thermal energy is transferred from large to small scales, resulting enclosed in a more thin toroid.

\bigskip

\section{\bf Formulation of the problem}

At this stage of the analysis, the problem for determining $f$ and $f_\theta$ 
is formulated through the closed von K\'arm\'an--Howarth and Corrsin equations, 
which are here reported
\bea
\ds \frac{\partial f}{\partial t} = 
\ds  \frac{K(r)}{u^2} +
\ds 2 \nu  \left(  \frac{\partial^2 f} {\partial r^2} +
\ds \frac{4}{r} \frac{\partial f}{\partial r}  \right) +\frac{10 \nu}{\lambda_T^2} f 
\label{vk-h II}
\eea
\bea
\ds  \frac{\partial  f_\theta}{\partial t}  =
\ds  \frac{G(r)}{\theta^2} 
+2 \chi  \left( \frac{\partial^2 f_\theta}{\partial r^2}  
+ \frac{2}{r} \frac{\partial f_\theta}{\partial r} \right) 
 +  \frac{12 \chi}{\lambda_\theta^2} f_\theta 
\label{f_theta II}
\eea
 where $K$ and $G$ are expressed by means of Eqs. (\ref{K closure}) and (\ref{G closure}) 
\bea
\begin{array}{l@{\hspace{-0.cm}}l}
\ds K= u^3 \sqrt{\frac{1-f}{2}} \ \frac{\partial f}{\partial r}, \\\\
\ds G= \theta^2 u  \sqrt{\frac{1-f}{2}} \ \frac{\partial f_\theta}{\partial r}
\end{array}
\eea
The boundary conditions of Eqs. (\ref{vk-h II}) and (\ref{f_theta II}) are 
\bea
\begin{array}{l@{\hspace{-0.cm}}l}
\ds f (0) = 1, \ \ \ \ \lim_{r \rightarrow \infty} f(r) = 0, \\\\
\ds f_\theta (0) = 1, \ \ \ \ \lim_{r \rightarrow \infty} f_\theta(r) = 0
\end{array}
\label{bc}
\eea
As $\chi$ and $\nu$ are considered to be assigned quantities, according to  
Eqs. (\ref{vk-h II}) and (\ref{f_theta II}), the solutions $f_\theta$ are  related to $f$, whereas  
$f$ will not depend upon $f_\theta$.
The energy spectrum $E(\kappa)$ and the transfer function $T(\kappa)$ are the Fourier transforms of $f$ and $K$ (\cite{Batchelor53})
\bea
\left[\begin{array}{c}
\ds E(\kappa) \\\\
\ds T(\kappa)
\end{array}\right]  
= 
 \frac{1}{\pi} 
 \int_0^{\infty} 
\left[\begin{array}{c}
 \ds  u^2 f(r) \\\\
 \ds K(r)
\end{array}\right]  \kappa^2 r^2 
\left( \frac{\sin \kappa r }{\kappa r} - \cos \kappa r  \right) d r 
\label{Ek}
\eea
as well as the temperature spectrum $\Theta(\kappa)$ 
and the temperature transfer function $\Gamma(\kappa)$ are calculated as 
(\cite{Ogura})
\bea
\left[\begin{array}{c}
\ds \Theta(\kappa) \\\\
\ds \Gamma(\kappa)
\end{array}\right]  
= 
 \frac{2}{\pi} 
 \int_0^{\infty} 
\left[\begin{array}{c}
 \ds  \theta^2 f_\theta(r) \\\\
 \ds G(r)
\end{array}\right]  
\kappa r \sin \kappa r \ dr 
\label{Tk}
\eea
in such a way that  
\bea
\begin{array}{l@{\hspace{-0.cm}}l}
\ds f_\theta(r) = \int_0^\infty \Theta (\kappa)  \frac{\sin \kappa r}{\kappa r} \ d\kappa, \ \ \ \
\ds G(r) = \int_0^\infty \Gamma (\kappa)  \frac{\sin \kappa r}{\kappa r} \ d\kappa
\label{Tk1}
\end{array}
\eea
and 
\bea
\int_0^\infty \Theta (\kappa)  \ d\kappa = \theta^2, \ \ \
\int_0^\infty \Gamma (\kappa)  \ d\kappa = 0
\label{Tk0}
\eea

\bigskip

\section{\bf Lyapunov analysis of temperature fluctuations}

This section presents the procedure for calculating the temperature fluctuations,
which is based on the Lyapunov analysis of the fluid strain (\cite{deDivitiis_1}), and on Eq. (\ref{P}).

In order to obtain the temperature fluctuation, consider now the relative motion between two contiguous particles, expressed by the infinitesimal separation vector $d{\bf x}$ which obeys to the equation
\bea
d \dot{\bf x} = \nabla {\bf u} \ d {\bf x}
\label{dx}
\eea 
where the velocity gradient follows the Navier--Stokes equations.
As observed by \cite{deDivitiis_1}, in turbulence, $d{\bf x}$ is much faster than the fluid state variables, and the Lyapunov analysis of Eq. (\ref{dx}) provides the local deformation in terms of maximal Lyapunov exponent $\Lambda \equiv \lambda(0) >0$
\bea
\frac{\partial {\bf x}}{\partial {\bf x}_0} \approx {\mbox e}^{\Lambda (t - t_0)} 
\label{stretch}
\eea
Now, the map $\bfchi$ : ${\bf x}_0 \rightarrow {\bf x}$, 
is the function which determines the current position $\bf x$ of a fluid particle located at the referential position ${\bf x}_0$ at $t = t_0$ (\cite{Truesdell77}).
Equation (\ref{P}) can be written in terms of the referential position 
${\bf x}_0$
\bea
\ds \frac{\partial  \vartheta}{\partial t}  =  \left( -\frac{\partial  \vartheta}{\partial x_{0 p}} u_h +
\chi
 \frac{\partial^2 \vartheta}  {\partial x_{0 p} \partial x_{0 q}} 
\frac{\partial x_{0 q}}{\partial x_{h}} \right)  \ \frac{\partial x_{0 p}}{\partial x_{h}} 
\label{N-Sr}
\eea
The adoption of the referential coordinates allows to factorize the temperature fluctuation and to express it in the Lyapunov exponential form of the local deformation. As this deformation is assumed to be much more rapid than $\ds {\partial  \vartheta}/{\partial x_{0 p}} u_h$ and 
$ \chi  {\partial^2 \vartheta}/  {\partial x_{0 p} \partial x_{0 q}}$,
the temperature fluctuation can be obtained integrating  Eq. (\ref{N-Sr}) with respect to the time, where
$\ds {\partial  \vartheta}/{\partial x_{0 p}} u_h$ and $ \chi  {\partial^2 \vartheta}/  {\partial x_{0 p} \partial x_{0 q}}$ are considered to be constant 
\bea
\ds   \vartheta \approx   \frac{1}{\Lambda} \left( -\frac{\partial  \vartheta}{\partial x_{0 p}} u_h +
\chi
 \frac{\partial^2 \vartheta}  {\partial x_{0 p} \partial x_{0 q}} 
\right)_{t = t_0}   \  \approx 
\frac{1}{\Lambda} \left(  \frac{\partial \vartheta} {\partial t} \right)_{t = t_0} 
\label{fluc_v2_0}
\label{fluc_v2}
\eea
This assumption is justified by the fact that, according to the classical formulation
of motion of continuum media (\cite{Truesdell77}),  
$\ds {\partial  \vartheta}/{\partial x_{0 p}} u_h$ and 
$\chi  {\partial^2 \vartheta}/  {\partial x_{0 p} \partial x_{0 q}}$ are smooth functions of $t$ --at least during the period of a fluctuation--  whereas the fluid deformation varies very rapidly according to Eqs. (\ref{dx})-(\ref{stretch}).

{\color{black} For what concerns the temperature gradient, its evolution is described by an equation arising from Eq. (\ref{fluc_v2_0})
\bea
\ds \frac{\partial}{\partial t} \nabla \vartheta \approx
\Lambda \nabla \vartheta 
\label{nabla T}
\eea
Although $\Lambda >0$ determines the exponential growth of $\nabla \vartheta$,
$\Lambda \approx \sqrt{\langle \vert \vert \nabla {\bf u} \vert \vert^2 \rangle}$ is a bounded quantity and this excludes that $\nabla \vartheta$ can diverge in finite time.
This is in line with \cite{Tran08}, where the temperature gradient is shown to be bounded due to the smoothness of $\nabla {\bf u}$.}

\bigskip

\section{\bf Statistical analysis of temperature derivative}

As explained in this section, the Lyapunov analysis of the local deformation
and some plausible assumptions about the statistics of $\bf u$ and $\vartheta$
lead to the calculation of the distribution function of $\partial \vartheta / \partial r$ 
and of all its dimensionless statistical moments.

The statistical properties of $\partial \vartheta / \partial r$, are here investigated 
expressing velocity and temperature through the following canonical decomposition 
(\cite{Ventsel, deDivitiis_3})
\bea
{\bf u} = \Sum_k  \hat{\bf U}_k \xi_k, \ \ \ \vartheta = \Sum_k  \hat{\Theta}_k \xi_k
\label{X0}
\eea 
where $\hat{\bf U}_k$ and $\hat{\Theta}_k$ are proper coordinate functions of $t$ and $\bf x$, and $\xi_k$ ($k=1, 2, ...$) are dimensionless independent stochastic variables which satisfy
\bea
\left\langle \xi_k \right\rangle = 0, \ \ 
\left\langle \xi_i \xi_j \right\rangle = \delta_{i j}, \ \ 
\left\langle \xi_i \xi_j \xi_k  \right\rangle =  \varpi_{i j k} \ p, \ \ 
 \vert p \vert >>> 1, \ \ 
\left\langle \xi_k^4 \right\rangle = O(1)
\label{X1}
\eea
where $\varpi_{i j k}$ = 1 for $i = j = k$, else $\varpi_{i j k}$=0.
According to \cite{deDivitiis_3}, the variables $\xi_k$ are chosen in such a way
that $\vert \langle \xi_k^3  \rangle \vert >>> 1$, $k=1, 2, ...$ so that  $\xi_k$ can describe the mechanism 
of thermal and kinetic energy cascade.

The dimensionless temperature fluctuation $\hat{\vartheta}$ is obtained
in terms of $\xi_k$ substituting Eq. (\ref{X0}) into Eq. (\ref{fluc_v2})
\bea
\hat{\vartheta} =  \Sum_{i j} A_{i j} \xi_i \xi_j 
+ \frac{1}{Pe} \Sum_{k} b_k \xi_k  
\label{fluc theta}
\eea
where $r = \hat{r} \lambda_T$, $\vartheta = \hat{\vartheta} \ \theta$, whereas 
$Pe = R \ Pr$ and $R= u \lambda_T/ \nu$ are P\'eclet and Reynolds numbers referred  to the Taylor scale, and $Pr = \nu/\chi$ is the fluid Prandtl number, therefore
$\Sum_{i j} A_{i j} \xi_i \xi_j$ and $1/Pe \Sum_{k} b_k \xi_k$ arise from convective term and fluid conduction, respectively. 
Thanks to the isotropy, both $\bf u$ and $\vartheta$ are two gaussian stochastic variables (\cite{Ventsel, Lehmann99}), accordingly, Eq. (\ref{fluc theta}) satisfies the Lindeberg condition, a very general necessary and sufficient condition for satisfying the central limit theorem (\cite{Ventsel, Lehmann99}). This condition does not apply to 
$\ds \partial \vartheta /\partial r \equiv \lim_{r \rightarrow 0} \Delta \vartheta/r$.
In fact, as $\Delta \vartheta$ is the difference between two correlated gaussian variables, its PDF could be a non--gaussian distribution function.

Now, to obtain the PDF of the dimensionless temperature gradient, the fluctuation $\ds \partial \hat{\vartheta} /\partial \hat{r}$ is first calculated in function of $\xi_k$
\bea
\frac{\partial \hat{\vartheta}}{\partial \hat{r}} 
 =   \Sum_{i j} \frac{\partial A_{i j}}{\partial \hat{r}}  \xi_i \xi_j 
+ \frac{1}{Pe}  \Sum_{k} \frac{\partial b_k}{\partial \hat{r}} \xi_k 
    \equiv L + S + P + N
\label{fluc T}
\eea
This fluctuation consists of the contributions $L$, $S$, $P$ and $N$, appearing into Eq. (\ref{fluc T}):
in particular, $L$ is the sum of all linear terms due to the fluid conductivity, $S \equiv S_{i j} \xi_i \xi_j$ is the sum of all semidefinite bilinear forms arising from the convective term, whereas $P$ and $N$  are, respectively, the sums of definite positive and negative quadratic forms, which derive from the convective term.
The quantity $L+S$ tends to a gaussian random variable being the sum of statistically orthogonal terms, while $P$ and $N$ do not, as they are linear combinations of squares (\cite{Madow40, Lehmann99}).
Their general expressions are  
$
 P = P_0 + \eta_1  +  \eta_2^2,  \ \
 N = N_0 + \zeta_1 -  \zeta_2^2  
$,
where $P_0$ and $N_0$ are constants, and $\eta_1$, $\eta_2$, $\zeta_1$ and  $\zeta_2$ are four different centered random gaussian variables which are mutually uncorrelated thanks to the hypotheses of fully developed chaos and isotropy. 
Therefore, the longitudinal fluctuation of the temperature derivative can be written as
\bea
\begin{array}{l@{\hspace{+0.2cm}}l}
\ds \frac {\partial \hat{\vartheta}}{\partial \hat{r}} = 
\psi_1 {\xi} + 
\ds \left( \psi_2  ( {\eta}^2-1 )  - \psi_3 ( {\zeta}^2-1 )  \right) 
\end{array}
\label{fluc3}
\eea
where $\xi$, ${\eta}$ and $\zeta$ are independent gaussian centered random variables
for which $\langle \xi^2 \rangle$ =
$\langle \eta^2 \rangle$ = 
$\langle \zeta^2 \rangle$=1, 
and $\psi_1$ $\psi_2$ and $\psi_3$ are given quantities. 
Due to the isotropy, the skewness of ${\partial \hat{\vartheta}}/{\partial r}$ must be equal to zero, thus $\psi_2 = \psi_3$, and 
\bea
\begin{array}{l@{\hspace{+0.2cm}}l}
\ds \frac {\partial \hat{\vartheta}}{\partial \hat{r}} = 
\psi_1 {\xi} + \psi_2  
\ds \left(   {\eta}^2   -  {\zeta}^2  \right) 
\end{array}
\label{fluc3A}
\eea
Furthermore, comparing the terms of  Eqs. (\ref{fluc3A}) and (\ref{fluc T}),
we obtain that $\psi_1$ and $\psi_2$ are related with each other and that their ratio
$\psi = \psi_1/ \psi_2$ depends on the P\'eclet number 
\bea
\frac{4 \psi_2^2 }{ \psi_1^2} =
\frac{ \left\langle (P+N)^2 \right\rangle}{\left\langle ( S_{i j} \xi_i \xi_j + 1/Pe \ \partial b_k/\partial \hat{r} \xi_k )^2 \right\rangle}
\label{psi_ratio}
\eea
Taking into account the properties (\ref{X1}) of $\xi_k$ ($\langle \xi_k^3 \rangle >>> 1$, 
$\langle \xi_k^4 \rangle =O(1)$), and in view of Eq. (\ref{psi_ratio}), we found  
\bea
\psi \equiv \frac{\psi_2}{ \psi_1} = C \sqrt{Pe}
\eea 
where $C$ is a proper constant which has to be identified.
Hence, the dimensionless longitudinal temperature derivative is 
\bea
\begin{array}{l@{\hspace{+0.2cm}}l}
\ds \frac {\vartheta_r}{\sqrt{\langle \vartheta_r^2} \rangle} =
\ds \frac{   {\xi} + \psi \left( {\eta}^2 - {\zeta}^2 \right) }
{\sqrt{1+4 \psi^2} } 
\end{array}
\label{Tfluc4}
\eea 
In order to identify $C$, observe that Eq. (\ref{Tfluc4}) is formally similar to the expression of the longitudinal velocity derivative ${\partial u_r/\partial r}$ obtained in \cite{deDivitiis_1} 
(see also the appendix)
\bea
\begin{array}{l@{\hspace{+0.2cm}}l}
\ds \frac {\partial u_r/\partial r}{\sqrt{\langle (\partial u_r/\partial r)^2} \rangle} =
\ds \frac{   {\xi_u} + \psi_u \left( \chi ( {\eta_u}^2-1 )  -  
\ds  ( {\zeta_u}^2-1 )  \right) }
{\sqrt{1+2  \psi_u^2 \left( 1+ \chi^2 \right)} } 
\end{array}
\label{fluc4 bis}
\eea 
$\xi_u$, $\eta_u$ and $\zeta_u$ are independent centered gaussian random variables
with $\langle \xi_u^2 \rangle = \langle \eta_u^2 \rangle = \langle \zeta_u^2 \rangle$ =1,
and
\bea
\psi_u(R) =  
\sqrt{\frac{R}{15 \sqrt{15}}} \
\hat{\psi}_u(0), \ \ \ \ \hat{\psi}_u(0) = O(1), \ \ \ \chi=\chi(R) = O(1)  
\label{Rl bis}
\eea
$\chi \ne 1$ provides a negative skewness of ${\partial u_r/\partial r}$,
whereas $\hat{\psi}_u(0)\simeq 1.075$ is determined through an approximate estimation 
of the critical value of $R$ (\cite{deDivitiis_1}).
Now, when $Pr =$ 1,  
it is reasonable to assume that the ratio between linear and quadratic terms of Eq. (\ref{Tfluc4}) is equal to that of the corresponding terms of Eq. (\ref{fluc4 bis}). 
Accordingly, $\psi \simeq \psi_u$ and this identifies an approximate value of $C$
\bea
\ds C \approx \frac{\hat{\psi}_u(0)}{15^{3/4}} \
 \simeq 0.141
\eea  

The distribution function of the temperature derivatives is thus
expressed through the Frobenius--Perron equation
\bea
\begin{array}{l@{\hspace{+0.0cm}}l}
F(\vartheta_r') = \hspace{-0.mm}
\ds \int_\xi \hspace{-0.mm}
\int_\eta  \hspace{-0.mm}
\int_\zeta \hspace{-0.mm}
p(\xi) p(\eta) p(\zeta) \
\delta \left( \vartheta_r'\hspace{-0.mm}-\hspace{-0.mm}\vartheta_r(\xi, \eta ,\zeta) \right)
d \xi \ d \eta \ d \zeta
\end{array}
\label{Tfrobenious_perron}
\eea 
where $\vartheta_r(\xi, \eta ,\zeta)$ is determined by Eq. (\ref{Tfluc4}), $\delta$ is the Dirac delta and $p$ is a centered gaussian PDF with standard deviation equal to one.

Finally, the dimensionless statistical moments of $\vartheta_r$
are easily calculated considering that $\xi$, $\eta$ and 
$\zeta$ are independent gaussian variables
\bea
\begin{array}{l@{\hspace{+0.2cm}}l}
\ds H_n \equiv \frac{\left\langle \vartheta_r^n \right\rangle}
{\left\langle \vartheta_r^2\right\rangle^{n/2} }
= 
\ds \frac{1} {(1+4  \psi^2)^{n/2}} 
\ds \sum_{k=0}^n 
\left(\begin{array}{c}
n  \\
k
\end{array}\right)  \psi^k
 \langle \xi^{n-k} \rangle 
  \langle (\eta^2  - \zeta^2 )^k \rangle 
\end{array}
\label{Tm1}
\eea

It is worth remarking that, for non--isotropic turbulence or in more complex cases with boundary conditions, the stochastic variables $\xi_k$ could not satisfy the Lindeberg condition, thus $\vartheta$ will be not distributed following a Gaussian PDF, and Eq. (\ref{Tfluc4}) changes its analytical form and can incorporate more intermittent terms (\cite{Lehmann99}) which give the deviation with respect to the isotropic turbulence. Hence, the absolute statistical moments of $\vartheta_r$ will be greater than those calculated with Eq.(\ref{Tm1}), indicating that, in a more complex situation than the isotropic turbulence, the intermittency of $\vartheta_r$ can be significantly stronger.

\bigskip

\section{\bf Self--Similar temperature spectrum}

An ordinary differential equation which describes the spatial evolution of
$f_\theta$ is now derived from Eq. (\ref{f_theta II}), adopting the hypothesis 
of self--similarity of \cite{Karman49}, \cite{George1, George2},
and using the proposed closure of the Corrsin equation. 

The idea of self--preserving correlation function consists in what follows:
far from the initial condition, the simultaneous effect of thermal and kinetic energy 
cascade with the fluid conductivity and viscosity acts keeping $f_\theta$ similar in the time. 
According to \cite{George1, George2}, $f_\theta$ can be scaled with respect to $\lambda_T(t)$, thus 
\bea
f_\theta = f_\theta (\hat{r}), \ \mbox{where} \ \ \   \hat{r} = \frac{r}{\lambda_T(t)} 
\label{similar}
\eea
Substituting Eq. (\ref{similar}) into Eq. (\ref{f_theta II}), we obtain
\bea
 -
\ds \frac{d f_\theta}{d \hat{r}} \ \frac{\hat{r}}{u} \frac{d \lambda_T}{dt}  = 
\ds  \sqrt{\frac{1-f}{2}} \ \frac{d f_\theta}{d \hat{r}}+
\ds  \frac{2}{R \ Pr}  \left(  \frac{d^2 f_\theta } {d \hat{r}^2} +
\ds \frac{2}{\hat{r}} \frac{d f_\theta }{d \hat{r}}  \right) + \frac{12}{R \ Pr} 
\left( \frac{\lambda_T}{\lambda_\theta}\right)^2  f_\theta 
\label{f_theta II_adim}  
\eea
Therefore, the boundary problem given by Eqs. (\ref{f_theta II}) and (\ref{bc})
is here reduced to an ordinary differential equation of the second order in the variable $r$.
Equation (\ref{f_theta II_adim}) is a non--linear equation, and 
\bea
\begin{array}{l@{\hspace{-0.cm}}l}
\ds \frac{d u^2}{d t} = - \frac{10 \nu u^2}{\lambda_T^2}, \ \ \ \
\ds \frac{d \theta^2}{d t} = - \frac{12 k \theta^2}{\lambda_\theta^2}
\end{array}
\label{rate e T}
\eea
describe the evolution of kinetic and thermal energies.
Now, due to self--similarity, 
all the coefficients of Eq. (\ref{f_theta II_adim}) do not vary with the time 
(\cite{Karman38, Karman49}, \cite{George1, George2}), thus
\bea
\begin{array}{l@{\hspace{-0.cm}}l}
\ds R = \mbox{const}, \ \ \ \
\ds \frac{1}{u} \frac{d \lambda_T}{dt} = \mbox{const}, \ \ \ \
\ds \frac{\lambda_\theta}{\lambda_T} =  \mbox{const}
\end{array}
\label{coeff const}
\eea
As $\lambda_T$ follows Eq. (\ref{ult}) (see appendix), $\lambda_\theta$
is obtained from the constancy of $\lambda_\theta/\lambda_T$ 
\bea
\begin{array}{l@{\hspace{-0.cm}}l}
\ds \lambda_\theta(t) = \lambda_\theta(0) \sqrt{1 +10 \nu \ t /\lambda_T^2(0) }.
\end{array}
\label{l_theta(t)}
\eea
Thus, according to \cite{Warhaft} and  \cite{George1, George2}, the microscales $\lambda_T$, $\lambda_\theta$ and the rates $d\theta^2/dt$ and $d u^2/dt$ depend on the initial conditions of temperature and kinetic energy spectra.
Taking into account Eq. (\ref{ult}), we obtain  
\bea
\begin{array}{l@{\hspace{-0.cm}}l}
\ds \frac{1}{u} \frac{d \lambda_T}{dt} = \frac{5}{R} 
\end{array}
\eea
The self--similar solutions are searched over the whole range of $\hat{r}$, but for the dimensionless distances whose order magnitude exceeds $R$.
This corresponds to assume the self--similarity for all the frequencies of the energy spectrum, with the exception of the lowest ones (\cite{Karman38, Karman49}). 
Accordingly, $\partial f_\theta/\partial t$ can be neglected with respect to the other terms, thus $f_\theta (\hat{r})$ obeys to the following non--linear ordinary differential equation 
\bea
\begin{array}{l@{\hspace{-0.cm}}l}
\ds {R \ Pr}  \sqrt{\frac{1-f}{2}} \ \frac{d f_\theta}{d \hat{r}}+
\ds  {2}  \left(  \frac{d^2 f_\theta } {d \hat{r}^2} +
\ds \frac{2}{\hat{r}} \frac{d f_\theta }{d \hat{r}}  \right) + {12}  
\left( \frac{\lambda_T}{\lambda_\theta}\right)^2  f_\theta = 0
\end{array}
\label{f_theta II_adimv B}  
\eea
For $\hat{r} =0$, Eq. (\ref{f_theta II_adimv B}) gives
\bea 
\ds \frac{d^2 f_\theta (0)} {d \hat{r}^2} = -2 \left( \frac{\lambda_T}{\lambda_\theta}\right)^2
\label{ftheta20}
\eea
Hence, to calculate $f_\theta$, $\lambda_T /\lambda_\theta$ must be first specified into Eq. (\ref{f_theta II_adimv B}).
This is determined substituting Eq. (\ref{l_theta(t)}) into  Eq. (\ref{rate e T}) and integrating this latter from $t=0$ to $t$
\bea
\begin{array}{l@{\hspace{-0.cm}}l}
\ds \ln \left( \frac{\theta(t)}{\theta(0)}\right)  = 
{-\frac{3}{5} \left( \frac{\lambda_T}{\lambda_\theta}\right)^2  \frac{1}{Pr}} 
\ln \left( 1+ \frac{10 \nu}{\lambda_T(0)^2} t \right), \\\\
\ds \ln \left( \frac{u(t)}{u(0)}\right)  = 
-\frac{1}{2}  
\ln \left( 1+ \frac{10 \nu}{\lambda_T(0)^2} t \right)
\end{array}
\label{ln theta}
\label{ln u}
\eea
The fully self--similarity (mechanical and thermal) occurs when $\theta$
and $u$ are proportional to each other (\cite{George1, George2})
\bea
 \frac{\theta(t)}{\theta(0)}= \frac{u(t)}{u(0)} 
\eea
Hence, the ratio $\lambda_T /\lambda_\theta$ satisfying this condition
depends on  the Prandtl's number 
\bea
\frac{\lambda_\theta}{\lambda_T} = \sqrt{\frac{6}{5} \frac{1}{Pr}}
\label{l_theta/l_t 1}
\eea
Accordingly, $f''_\theta(0)$ is related to $Pr$ 
\bea
\ds \frac{d^2 f_\theta} {d \hat{r}^2} (0)= - \frac{5}{3} Pr
\label{l_theta/l_t 2}
\eea 
This self--similarity expresses a further link between $f$ and $f_\theta$.

\bigskip 

Observe that the solutions $f_\theta \in C^2 \left[0, \infty \right)$ of Eq. (\ref{f_theta II_adimv B}) with $d f_\theta/ d \hat{r}(0)=0$ and $Pr \ne 0$, tend to zero as $r \rightarrow \infty$, thus the boundary condition (\ref{bcT}) can be replaced by the following conditions in the origin
\bea
\ds  f_\theta (0) = 1, \ \
\ds \frac{d f_\theta} {d \hat{r}} (0)= 0
\label{bcT3}
\eea
Therefore, the boundary problem represented by Eqs. (\ref{f_theta II_adimv B}) and (\ref{bcT}), is reduced to the following initial condition problem written in the Cauchy's normal form
\bea
\begin{array}{l@{\hspace{+0.cm}}l}
\ds  \frac{d f_\theta}{d \hat{r}} =  F_\theta \\\\
\ds \frac{d F_\theta}{d\hat{r}} = -5 \ Pr  \ f_\theta -
\left( \frac{R \ Pr}{2} \sqrt{\frac{1-f}{2}}  + \frac{2}{\hat{r}} \right) F_\theta
\end{array}
\label{f_theta II_adimv C}  
\eea
the initial condition of which is 
\bea 
\ds f_\theta(0) = 1, \ \  F_\theta(0) = 0
\label{icT}
\eea

\bigskip

In conclusion, the self--similar functions $f$ and $f_\theta$ are calculated as the solutions of 
the ordinary differential system (\ref{vk-h2}) and (\ref{f_theta II_adimv C})
with the initial conditions (\ref{ic}) and (\ref{icT}).

\bigskip

\section{\bf Results and Discussion}

Temperature and velocity correlations are first calculated
with Eqs. (\ref{vk-h2}) and (\ref{f_theta II_adimv C}), for several values of $R$ and $Pr$. Thereafter, the statistical analysis of $\partial \vartheta / \partial r$ 
is carried out using the Lyapunov theory of the temperature derivative.

The case with $Pr \rightarrow 0$ is first considered.
This is a limit case of Eqs. (\ref{f_theta II_adimv C})--(\ref{icT}) corresponding to the following equation
\bea
\ds \frac{d^2 f_\theta}{d \hat{r}^2} + \frac{2}{\hat{r}} \frac{d f_\theta}{d \hat{r}}  = 0
\eea
which does not admit analytical solutions with the boundary conditions (\ref{bc}).

Conversely, when $Pr \rightarrow \infty$ ($\nu >>> \chi$),
\bea
\ds \lim_{\hat{r}\rightarrow 0} \frac{d^2 f_\theta}{d \hat{r}^2} = -\infty
\eea
and this gives the behavior of $f_\theta$ near the origin. For $\hat{r} >0$,  $f_\theta$ is obtained solving Eqs. (\ref{f_theta II_adimv C})--(\ref{icT}) by quadrature, in terms of $f$
\bea
\ln f_\theta (\hat{r}) = - \frac{10 \sqrt{2}}{R} \int_0^{\hat{r}} \frac{d \xi}{\sqrt{1-f(\xi)}}
\label{f_theta f}
\eea
This equation implies that, if $R$ is large enough and
\bea
f \simeq 1 - \left( \frac{r}{L_u}\right)^{2/3}, \ \ \mbox{then also} \ \
f_\theta \simeq 1 - \left( \frac{r}{L_\theta}\right)^{2/3}
\eea
where $L_\theta$= $L_u$ are length scales proportional to $\lambda_T$
\bea
L_u = L_\theta = \frac{R}{15 \sqrt{2}} \lambda_T, 
\eea
Therefore, if $Pr \rightarrow \infty$, the ratios between
the scales are
\bea
\frac{L_\theta}{L_u} = 1, \ \ \ \ \frac{\lambda_\theta}{\lambda_T} = 0
\eea

When $R$ and $Pr$ change, the ratios between the scales vary depending on the combined values of $R$ and $Pr$, therefore quite different situations occur.
\begin{figure}[h]
	\centering
\hspace{-45.mm}         \includegraphics[width=0.7\textwidth]{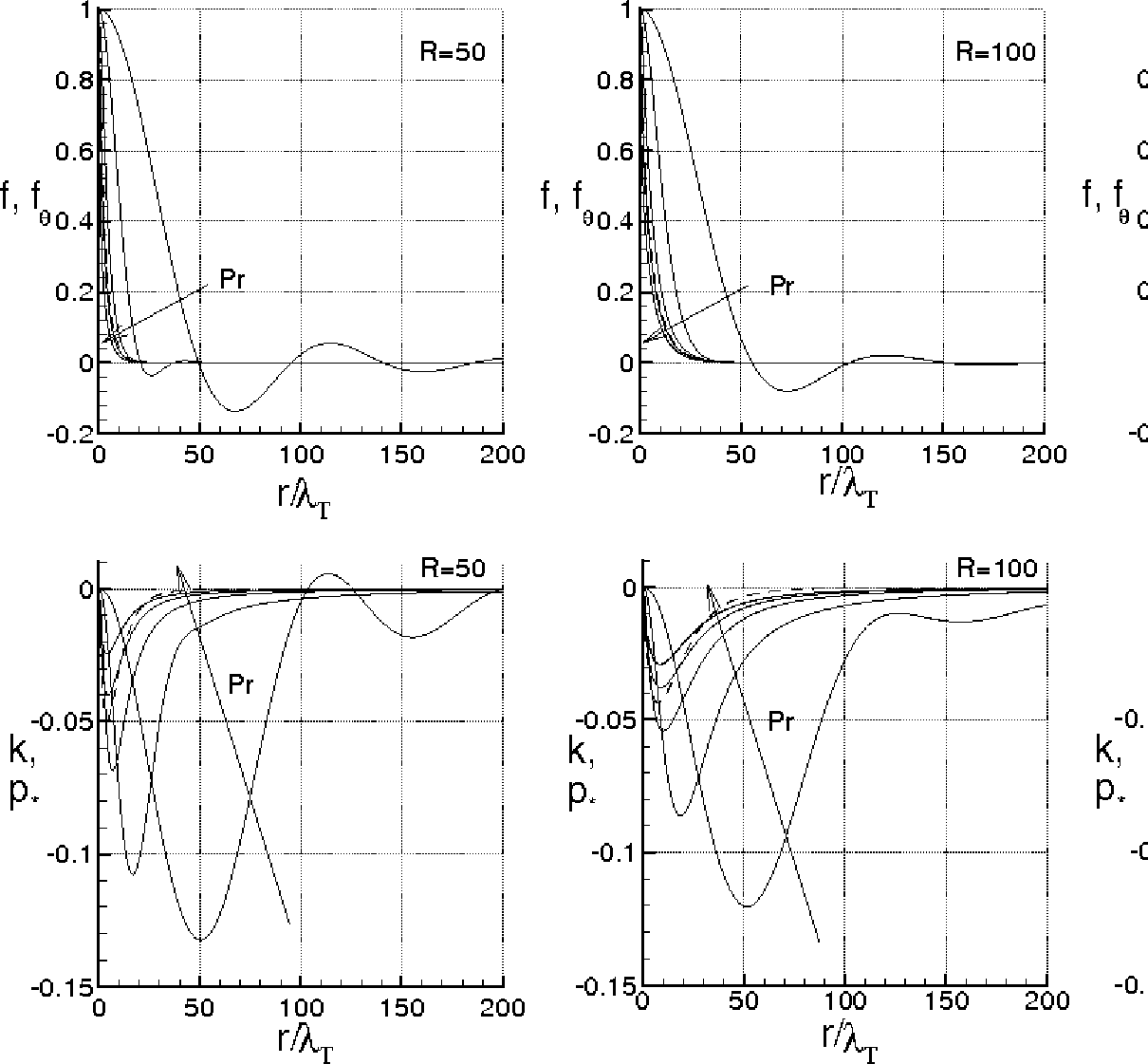}
\caption{Correlation functions for Pr= 10$^{-3}$, 10$^{-2}$, 0.1, 1.0 and 10, at different Reynolds numbers.
 Top: velocity correlation $f$ (dashed line) and temperature correlation $f_\theta$ (solid lines). Bottom: triple velocity correlation $k$ (dashed line) and triple velocity--temperature correlation $p_*$ (solid lines)}
\label{figura_1}
\end{figure}

In order to study the influence of $R$ and $Pr$ on $f_\theta$,  Eqs. (\ref{vk-h2}) and (\ref{f_theta II_adimv C}) are numerically solved for different values of $R$ and $Pr$.
The Reynolds number is assumed to be $R=$50,  100 and 300, whereas 
$Pr$ ranges from 0.001, to 10.
Figure \ref{figura_1} shows $f$ (dashed line) and $f_\theta$  (solid lines). 
The temperature correlation, related to $f$ through  Eq. (\ref{G closure}), is furthermore linked to $f$ by self--similarity (see Eq. (\ref{l_theta/l_t 1})). 
Therefore, when $R$ and $\lambda_T$ are given, the Corrsin microscale decreases 
as $Pr$ rises, and the curves of $f_\theta$ seem to collapse into a single diagram when $Pr \rightarrow \infty$. On the contrary, small values of $Pr$ determine large length scales of $f_\theta$ and $p_*$.
The case with $R$ = 50 is first considered. For $Pr$ =0.001, $f_\theta$ exhibits 
oscillations whose amplitude decreases as $\hat{r}$ rises.
When $Pr$ increases, oscillations magnitude and Corrsin microscale diminish, and for $0.01 < Pr < 0.1$ such oscillations vanish, being $f_\theta>$ 0.
The case $R=100$ differs from the previous one. The higher value of $R$ determines a sizable reduction of the oscillations, whereas the dimensionless scales of $f$ and $f_\theta$ are greater than the previous ones.
Next, for $R=300$, the dimensionless scales increase further, and $f_\theta>0$ is a monotonic function of $r$ for each value of $Pr$.

Accordingly, the triple correlation $p_*$ varies with $R$ and $Pr$.
For $R=50$, small values of $Pr$ (0.001) correspond to large dimensionless scales and sizable oscillations of $p_*$, whereas higher Prandtl numbers give reductions of these oscillations and of the length scales.
Increasing $R$ ($R$ = 100 and 300), the length scales rise, the oscillations disappear, and a reduction of $\vert p_*\vert_{MAX}$ is observed.
When $Pr$ = 0.7 and 1, $f_\theta \approx f$, and $p_* \approx k$, and this is in very good agreement with the classical experiments of \cite{Mills} which regards the turbulence behind heated grid.
\begin{figure}[h]
	\centering
 \hspace{-45.mm}        \includegraphics[width=0.70\textwidth]{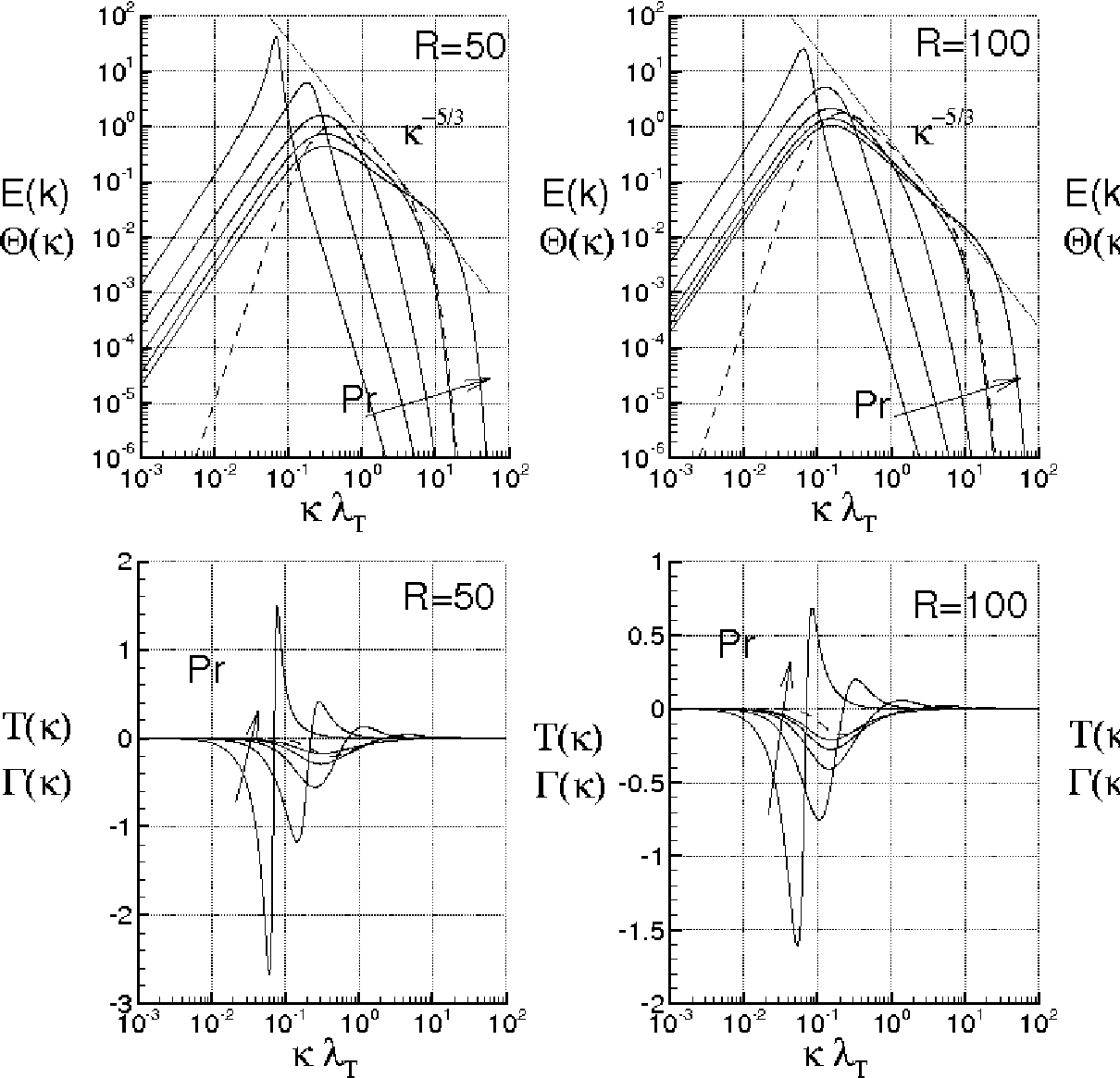}
\caption{Spectra for Pr= 10$^{-3}$, 10$^{-2}$, 0.1, 1.0 and 10, at different Reynolds numbers. 
Top: kinetic energy spectrum $E(\kappa)$ (dashed line) and temperature spectrum 
$\Theta(\kappa)$ (solid lines). Bottom: velocity transfer function $T(\kappa)$ (dashed line) and temperature transfer function $\Gamma(\kappa)$ (solid line)}
\label{figura_2}
\end{figure}

As far as $f(r)$ and $k(r)$ are concerned, these agree with the results of  \cite{deDivitiis_2}.

\begin{figure}[h]
\hspace{-0.mm}         \includegraphics[width=0.70\textwidth]{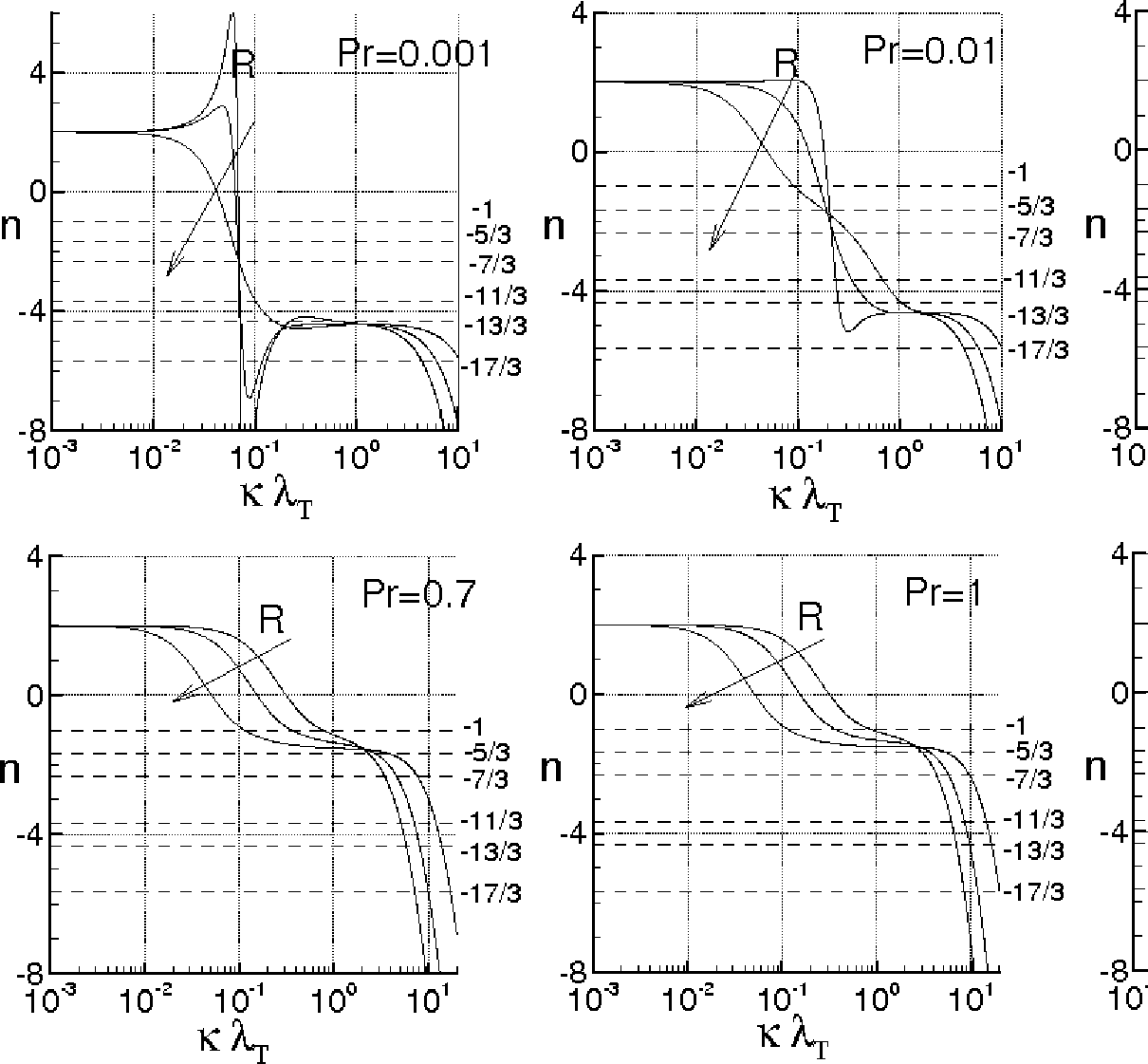}
\caption{Scaling exponent of the temperature spectrum calculated for Re = 50, 100 and 300, at different values of the Prandtl's number.}
\label{figura_3}
\end{figure}

The temperature spectra $\Theta(\kappa)$, calculated with Eq. (\ref{Ek}) and (\ref{Tk}), 
are depicted in  Fig. \ref{figura_2}.
The variations of $\Theta(\kappa)$ with $R$ and $Pr$ are quite peculiar and 
are consistent with the prior studies of the literature in the sense that there are
regions where $\Theta(\kappa)$ exhibits different scaling laws
$\Theta(\kappa) \approx \kappa^n$.
In any case, according to Eq. (\ref{G closure}), 
$n \rightarrow$ 2, as $\kappa \rightarrow$ 0.
For $Pr=$ 0.001, when $R$ ranges from 50 to 300, the temperature spectrum shows essentially two regions (see also Fig. \ref{figura_3}): one near the origin where 
$n \simeq 2$, and the other one, at higher values of $\kappa$, where $-17/3 < n < -11/3$,  (value very close to $-13/3$). 
The exponent $n$ varies rapidly at low Reynolds numbers,
whereas for higher $R$, $n$ exhibits more gradual variations.
The value of $n \approx -13/3$, here obtained in an interval around to 
$\hat{r} \approx$1, is in between the exponent proposed by \cite{Batchelor_3} ($-17/3$) and the value determined by \cite{Rogallo} ($-11/3$) with the numerical simulations.
Increasing $\kappa$, $n$ strongly diminishes, and $\Theta(\kappa)$ does not show scaling law.
With reference to Fig. \ref{figura_3}, when $Pr$=0.01, the three curves intersect with each other for $0.1<\hat{r}<1$, where $n \simeq -5/3$ and the curves exhibit inflection points.
Now, there is an interval near $\hat{r} \approx 1$ where $-17/3 < n < -13/3$, and this
is in agreement with \cite{Batchelor_3}.
For $Pr$ =0.1, the previous scaling law vanishes, whereas for $R=$ 50 and 100, $n$ changes with $\kappa$, and $\Theta(\kappa)$ does not show clear scaling laws.
When $R=300$, the birth of a small region is observed, where $n \approx -5/3$ has an inflection point.
For $Pr =$ 0.7 and 1, with $R =$ 300, the width of this region is increased, whereas
at $Pr$ = 10, and $R =$ 300, we observe two regions:
one interval where $n$ has a local minimum with $n \simeq -5/3$, and the other 
one where $n$ exhibits a relative maximum, with $n \simeq -1$.
For larger $\kappa$, $n$ diminishes and the scaling laws disappear.

Figure \ref{figura_2} reports also (on the bottom) the spectra 
$\Gamma(\kappa)$ (solid lines) and $T(\kappa)$ (dashed lines) which
describe the mechanism of kinetic and thermal energy cascade. 
As these latter do not modify the value of $\theta$ and $u$, 
$\int_0^{\infty } T(\kappa) d \kappa \equiv$ 0, and
 $\int_0^{\infty }\Gamma(\kappa) d \kappa \equiv$ 0.

\bigskip

\begin{figure}[h]
\hspace{-0.mm}         \includegraphics[width=0.70\textwidth]{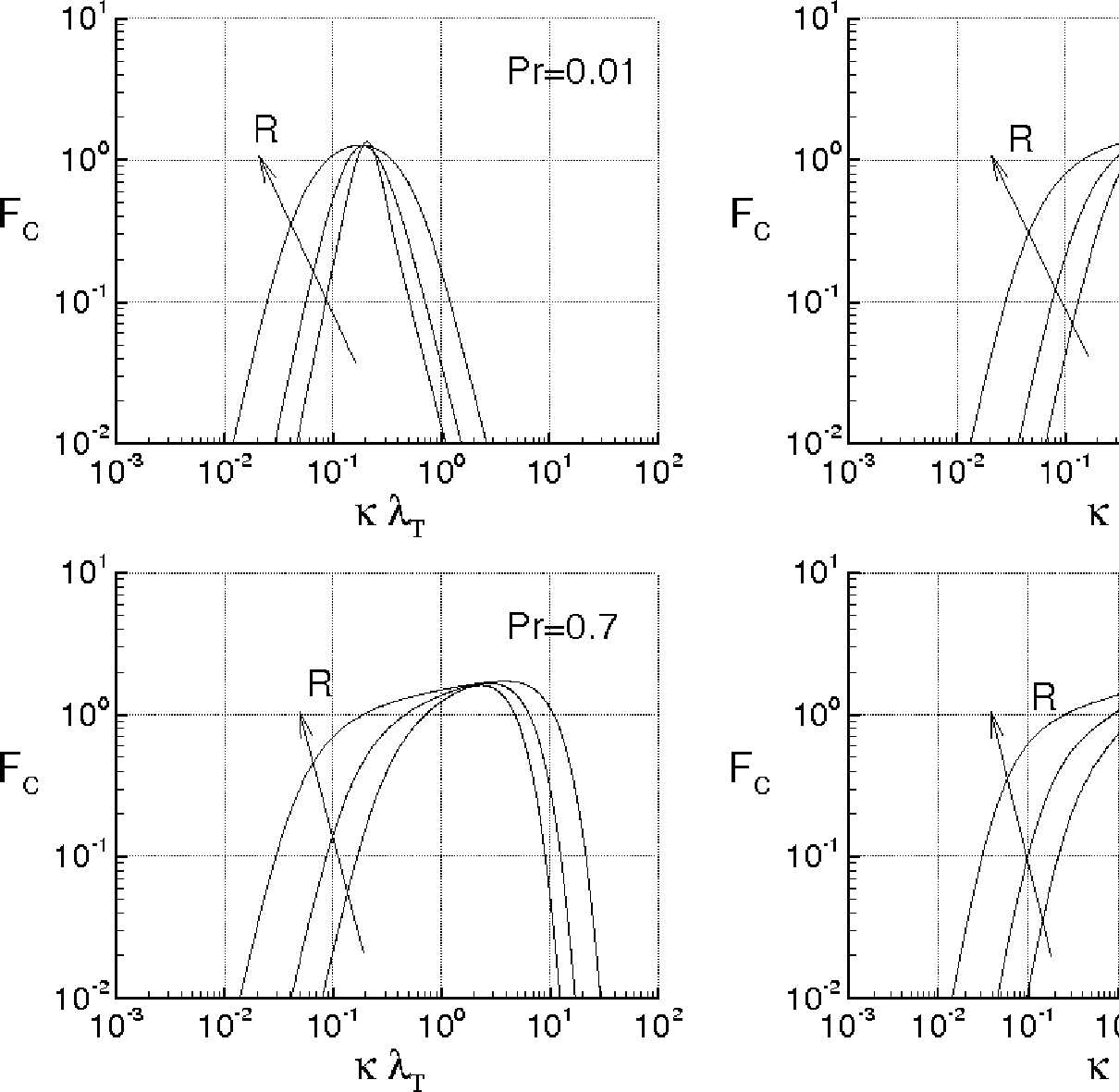}
\caption{Corrsin function for R= 50, 100 and 300, at different values of Prandtl's numbers.}
\label{figura_4}
\end{figure}

The presence of the scaling law $n \simeq -5/3$ agrees with the theoretical
arguments of \cite{Corrsin_2, Obukhov} (see also \cite{Mydlarski, Donzis} and references therein).
For large values of $R$ and $Pr$, $\Theta(\kappa)$ behaves like 
\bea
\Theta(\kappa) = C_\theta \epsilon^{-1/3} \epsilon_\theta \kappa^{-5/3}, 
\eea
in the inertial--convective range, where
\bea
\epsilon_\theta = 12 \chi \frac{\theta^2}{\lambda_\theta^2}, \ \ \ \
\epsilon = 15 \nu \frac{u^2}{\lambda_T^2}
\eea
and $C_\theta$ is the so--called Corrsin--Obukhov constant, a quantity 
of the order of the unity. This study identifies $C_\theta$ by means of the obtained results. 
First, the following quantity 
\bea
F_C(\kappa)= \Theta(\kappa) \epsilon^{1/3} \kappa^{5/3} \epsilon_\theta^{-1}
\eea
--here called Corrsin function-- is calculated in terms of $\kappa$, $Re$ and $Pr$.
Thereafter, $C_\theta$ is determined in such a way that $C_\theta = F_C(\kappa)$ in the range where $F_C(\kappa)$ is about constant.
A different definition of the Corrsin--Obukhov constant $C_{\theta 1}$ 
can be made with respect to  the one dimensional spectrum
\bea
\ds \frac{d \Theta_1} {d \kappa_1} (\kappa_1) = -\frac{\Theta(\kappa_1)}{2 \kappa_1}, \   
\mbox{where}  \
\Theta_1(\kappa) = C_{\theta 1} \epsilon^{-1/3} \epsilon_\theta \kappa^{-5/3}
\ \ \mbox {and}  \ C_{\theta 1}  = 0.3 C_{\theta}
\label{1D spectrum}
\eea
Figure \ref{figura_4} reports $F_C$ in terms of $\kappa$ for different
values of $R$ and $Pr$. For $Pr$ =0.01, at the relatively small Reynolds number,
the temperature spectrum does not follow the law $\kappa^{-5/3}$, thus $C_\theta$
is not defined, whereas at $R$ = 300 the diagram shows a region with a local maximum
where the variations of $F_C$ are relatively small. This maximum identifies
$C_\theta$ which results to be about 1.5 ($C_{\theta 1}\simeq 0.45$).
For $Pr$ =0.1, the larger scaling interval, implies a wider range where 
$F_C(\kappa) \simeq$ const, for each Reynolds number, resulting now
$C_\theta\simeq$ 1.8 ($C_{\theta 1}\simeq 0.54$).
When $Pr = O(1)$ (0.7 in the figure), the scaling law $\kappa^{-1}$, determines that 
$F_C$ slightly rises with $\kappa$, ranging from 1.4 to 2 ($C_{\theta 1} \simeq 0.42 \div 0.58$), values comparable with those obtained in  \cite{Mydlarski}.
For $Pr$ = 10, the region width where $\Theta \approx \kappa^{-1}$ increases,
thus $C_\theta$ is not defined.

\bigskip

The scaling law $k^{-1}$ is in line with the theoretical arguments proposed in \cite{Batchelor_2}, where
\bea
\Theta(\kappa) = C_B \sqrt{\frac{\nu}{\epsilon}} \ \epsilon_\theta \ \kappa^{-1}, 
\eea
in the viscous--convective range, where $C_B=O(1)$ is the Batchelor's constant. The present analysis identifies $C_B$ through the temperature spectra previously calculated. To this end, the following quantity
\bea
F_B(\kappa)= \Theta(\kappa) \kappa \sqrt{\frac{\epsilon}{\nu}} \epsilon_\theta^{-1}
\eea
here called Batchelor's function, is first calculated in function of $\kappa$, for several values of  $Re$ and $Pr$. The Batchelor's constant is estimated 
as $C_\theta = F_B(\kappa)$  in the region where $F_B(\kappa) \approx$ const 
(or at least exhibits a plateau). 
The Batchelor's constant $C_{B 1}$ can be defined with respect to  
the one dimensional spectrum (\ref{1D spectrum})  
\bea
\Theta_1(\kappa_1) = C_{B 1} \sqrt{\frac{\nu}{\epsilon}} \ \epsilon_\theta \ \kappa_1^{-1}, 
\ \ \mbox {where} \ \ C_{B 1}  = 0.5 C_{B}
\eea

\begin{figure}[h]
  \hspace{-0.mm}       \includegraphics[width=0.70\textwidth]{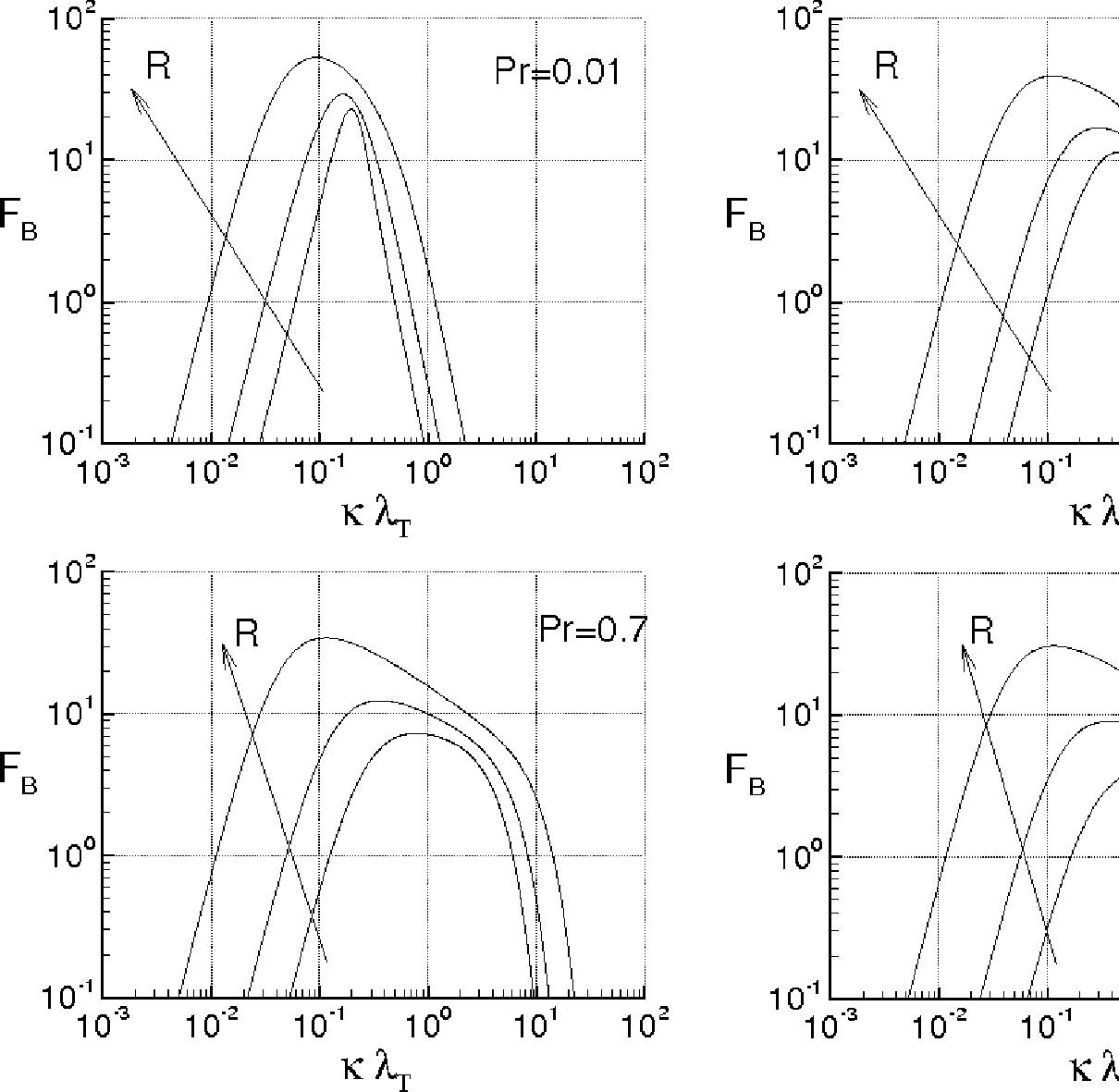}
\caption{Batchelor's function for R= 50, 100 and 300, at different values of Prandtl's numbers.}
\label{figura_5}
\end{figure}

In Fig.  \ref{figura_5}, $F_B$ is represented versus $\kappa$ for different values 
of $Re$ and $Pr$. It is apparent that $F_B \approx$ const when $Pr$ is high enough.
For $Pr$ =10, when $R$=  50, 100 and 300, 
the values of $C_B$ are about 5, 7 and 8, respectively
(that is $C_{B 1} \simeq$ 2.5, 3.5 and 4)
and this occurs for $1 <\kappa < 10$. 
These values are in quite good agreement with \cite{Donzis} (and references therein), and are consistent with the experiments of \cite{Grant} and \cite{Oakey} which deal with the temperature spectrum observed in the ocean.
\bigskip

\begin{figure}[ht]
	\centering
          \includegraphics[width=0.55 \textwidth]{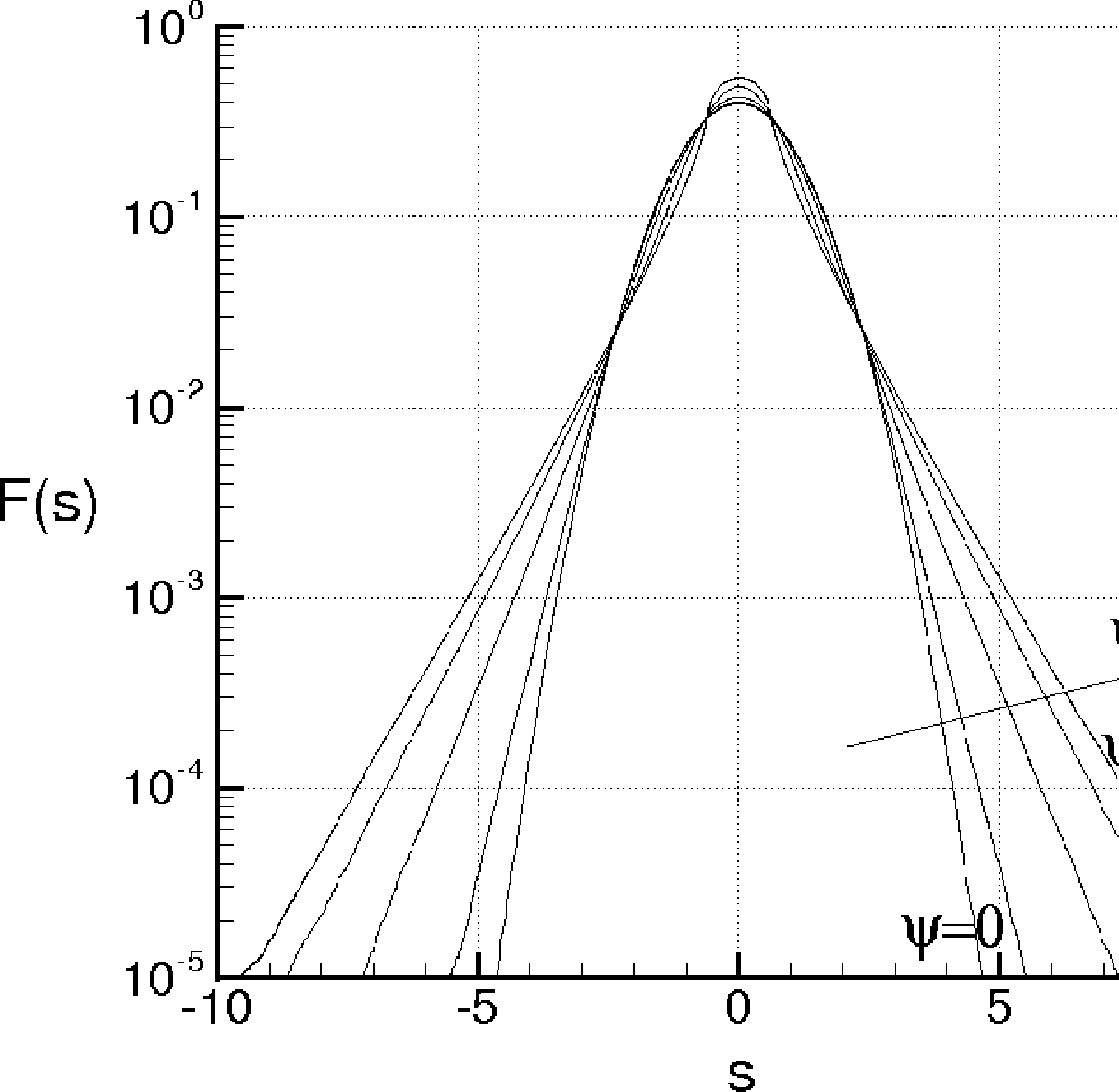}
\caption{Distribution function of the longitudinal temperature derivatives, at different values of $\psi=C \sqrt{Pr \ R}$}
\label{figura_6}
\end{figure}

\begin{figure}[ht]
	\centering
          \includegraphics[width=0.55 \textwidth]{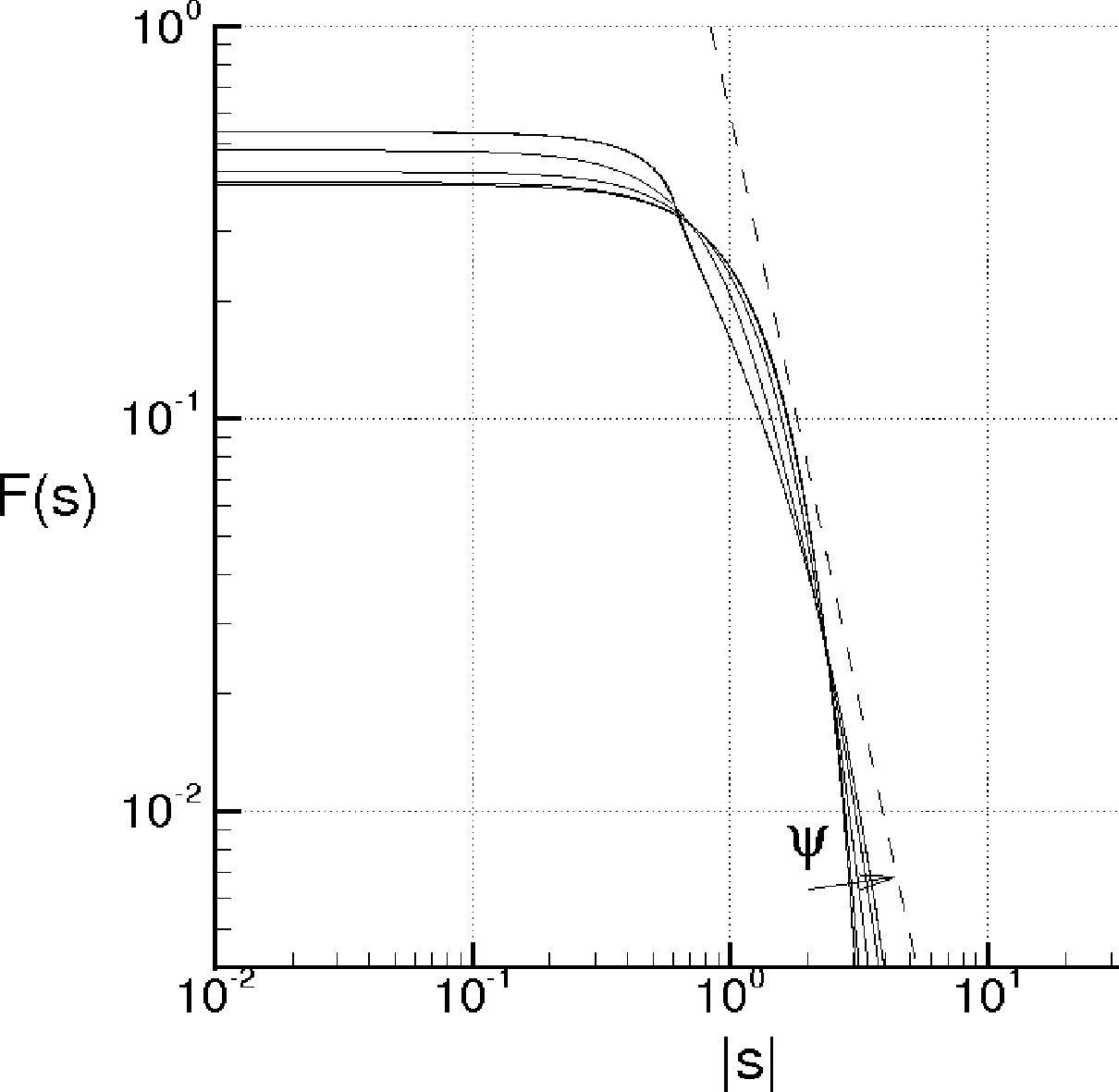}
\caption{\color{black} PDF in function of $\vert s \vert$}
\label{figura_6a}
\end{figure}

Next, in order to analyse the statistics of the temperature derivative, the  PDF
of $\partial \vartheta / \partial r $ is calculated with 
Eqs. (\ref{Tfrobenious_perron}) and (\ref{Tfluc4}), 
in function of the parameter $\psi = C \sqrt{Pr \ R}$.
This PDF is obtained with sequences of the variables $\xi$,  $\eta$ and $\zeta$, each generated by a gaussian random numbers generator. The distribution function is then calculated through the statistical elaboration of these data and Eq. (\ref{Tfluc4}). The corresponding results are shown in Fig. \ref{figura_6}, where the PDF is shown in terms of the dimensionless abscissa
\bea
s = \frac{\partial \vartheta /\partial r}{\sqrt{\left\langle \partial  \vartheta /\partial r^2 \right\rangle} }
\eea
These distribution functions are normalized, in order that their standard deviations are equal to the unity. These PDFs are even functions of $s$ and their tails rise with $\psi$ in such a way that the intermittency of 
$\partial \vartheta / \partial r$ increases with $\psi$, according to Eq. (\ref{Tm1}).
In the figure, the PDFs are calculated for $\psi=$ 0, 0.25, 0.5, 1., 10., $\infty$.
 For $\psi =$ 10 and $\infty$, the two curves are about overlapped.
{\color{black} More in detail, Fig. \ref{figura_6a} shows the PDFs in function of 
$\vert s \vert$ in logarithmic scale. For large P\'ecl\'et numbers, the PDF varies according to $s^{-m}$ (dashed line) with $m \simeq 3$ in the interval 
$2 \lesssim s \lesssim 4$, whereas for $s \gtrsim 4$ the exponent $m$ 
exhibits higher values. 
This is consistent with the results obtained by \cite{Fereday} which regard the PDF of the scalar concentration. It is worth remarking that the present results deal with the temperature derivative in isotropic turbulence, whereas the data of \cite{Fereday} concern the intermittency of the scalar concentration. Nevertheless, this comparison can be considered adequate, since, in isotropic turbulence, $\partial \vartheta/ \partial r$ exhibits an intermittency which varies with $Pe$, whereas $\vartheta$ remains a gaussian random variable.}

\begin{figure}[ht]
	\centering
         \includegraphics[width=0.55\textwidth]{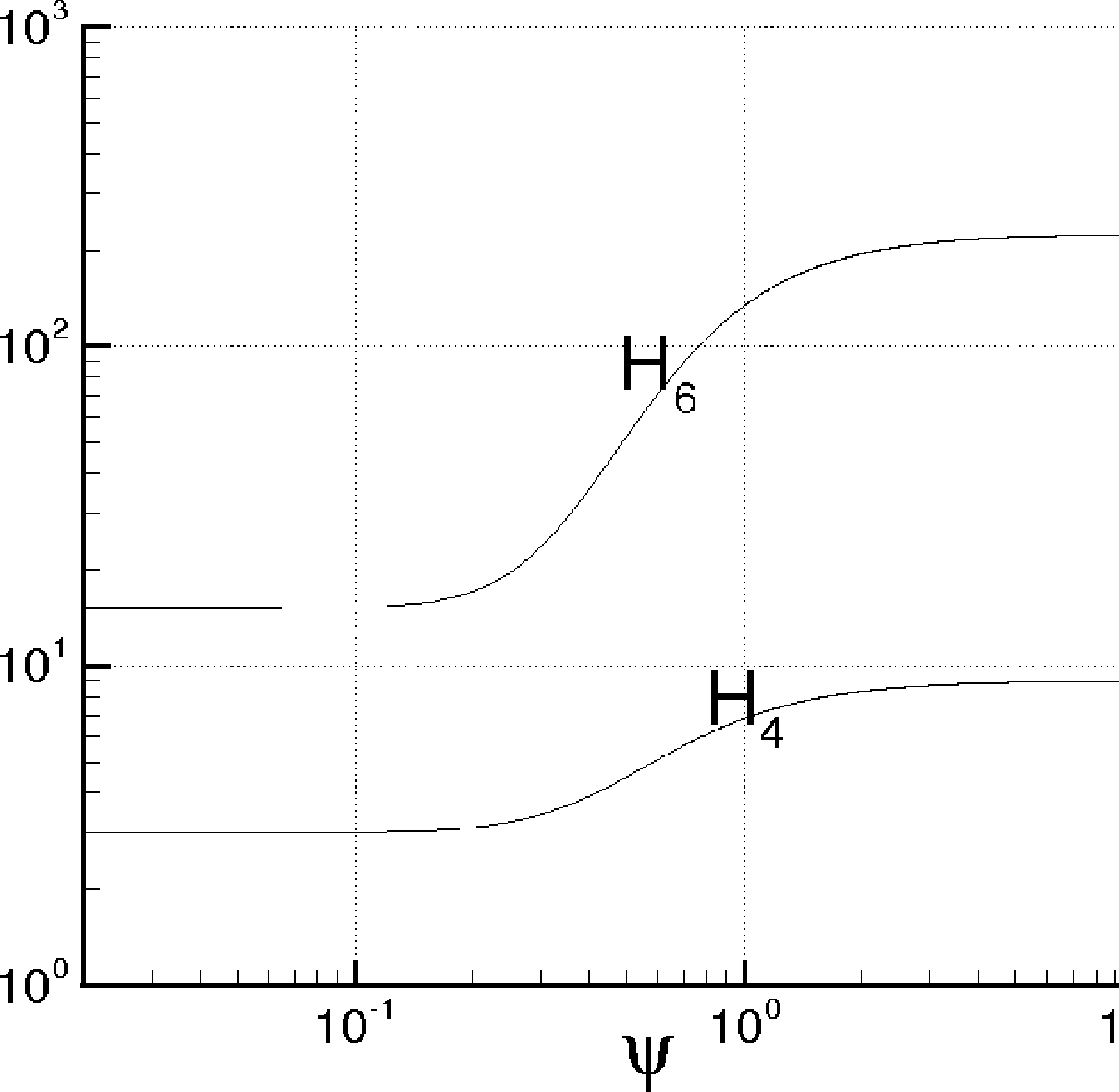}
\caption{Dimensionless statistical moments, $H_4$ and $H_6$ of 
$\partial \vartheta/\partial r$ in function of $\psi$.}
\label{figura_7}
\end{figure}

To study the intermittency, 
the flatness $H_4$ and the hyperflatness $H_6$, defined as
\bea
\ds H_4 = \frac{\langle s^4 \rangle}{ \langle s^2 \rangle^2}, \ \ \ \
\ds H_6 = \frac{\langle s^6 \rangle}{ \langle s^2 \rangle^3}
\eea
are shown in Fig. \ref{figura_7} in function of $\psi$.
For $\psi=$0, the PDF is gaussian, thus $H_4$ = 3 and $H_6$ = 15.
Increasing $\psi$, the non--linear terms $\eta$ and $\zeta$ determine an increment of $H_4$ and $H_6$, and  when $\psi \rightarrow \infty$ $H_4\rightarrow$ 9 and $H_6 \rightarrow$ 225.
These results are compared with the experiments of \cite{Sreenivasan}. 
In particular, the value of $C$ identified through the data of \cite{Sreenivasan} 
is $C\simeq 0.135$ against the value $C \approx 0.141$ here calculated. 

\begin{figure}[ht]
	\centering
         \includegraphics[width=0.55\textwidth]{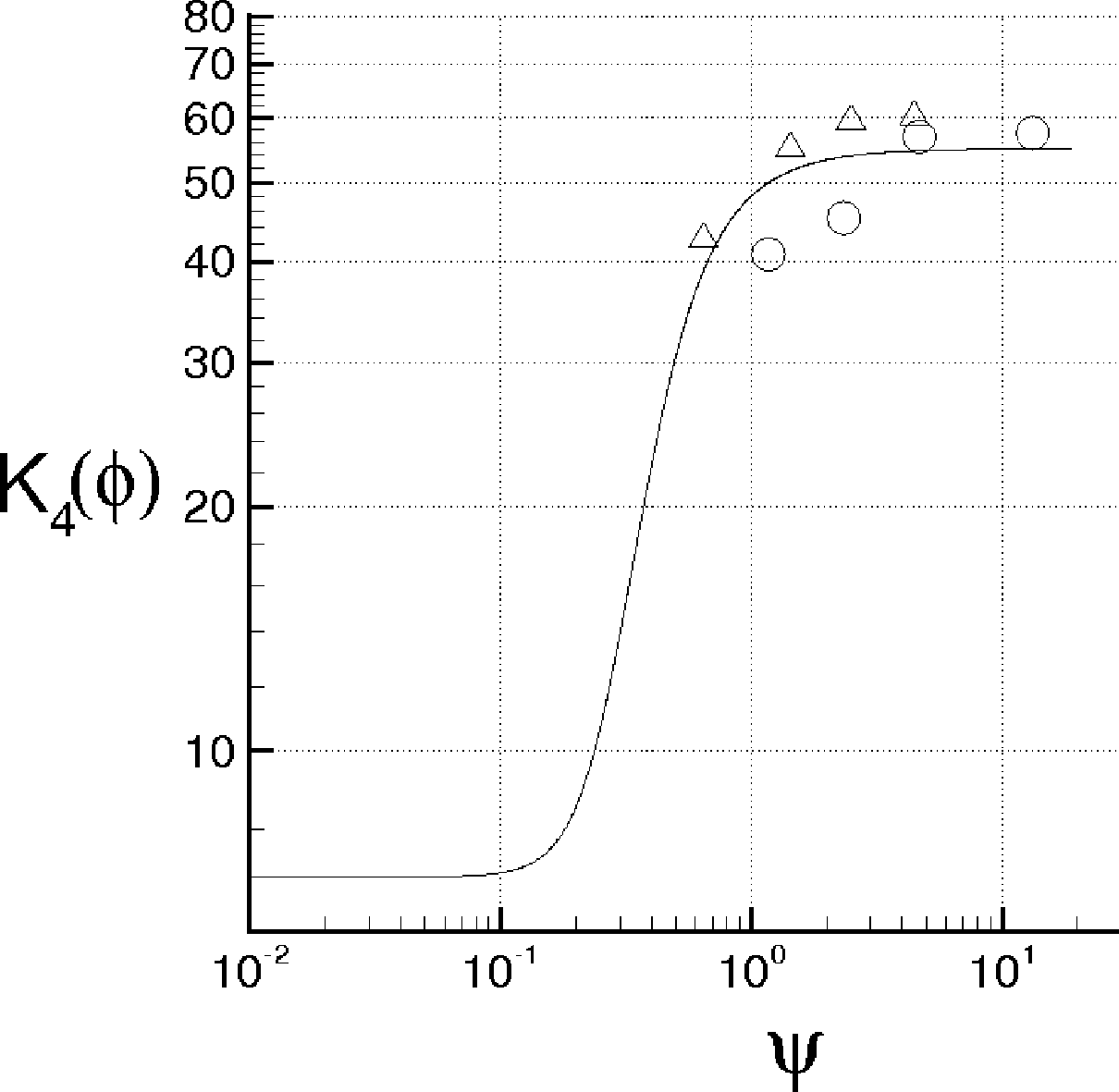}
\caption{\color{black} Comparison of the results: Kurtosis of temperature dissipation in function of $\psi$. The symbols represent the result by \cite{Burton}.}
\label{figura_8}
\end{figure}

{\color{black} Next, the statistics of the temperature dissipation  
\bea
 \varphi = \chi \nabla \vartheta \cdot \nabla \vartheta,
\eea
is analyzed in function of the Reynolds number. 
For this purpose, the Kurtosis of $\varphi$, $K_4(\varphi)$, is calculated with Eq.(\ref{Tfluc4}), taking into account that, due to isotropy, the three components of $\nabla \vartheta \equiv ( \vartheta_x, \vartheta_y, \vartheta_z)$ are identically distributed.
Furthermore, $\vartheta_x$, $\vartheta_y$ and $\vartheta_z$ are supposed to be statistically uncorrelated and this allows to analytically express  $K_4(\varphi)$ in terms of the dimensionless statistical moments of $\partial \vartheta/ \partial r$, according to
\bea
\ds K_4(\varphi)= \frac{H_8 - 4 H_6 +6 H_4 -3}{3(H_4^2+1-2 H_4)} + 2
\eea 
where $H_4$, $H_6$ and $H_8$ are calculated with Eq. (\ref{Tm1}).
Figure \ref{figura_8} shows $K_4(\varphi)$ in function of $\psi$,
and compares the values calculated with the present theory (solid line), with the kurtosis of the scalar energy dissipation obtained by \cite{Burton} through the nonlinear large--eddy simulations (symbols). The comparison shows that the data are in qualitatively good agreement.
More in detail, for $\psi \rightarrow \infty$, $K_4 =$ 55, whereas the results of \cite{Burton} give a value of about 60. This difference could be due to the fact that the
present analysis considers only the isotropic turbulence which tends to reduce the absolute dimensionless statistical moments of $\partial \vartheta/ \partial r$ and of $\varphi$, whereas the results of \cite{Burton} arise from the nonlinear large--eddy simulations. Next, in our calculation the components of $\nabla \vartheta$ are assumed to be statistically uncorrelated.
}

\bigskip

\section{\bf  Conclusions}

The finite--scale Lyapunov theory is adopted to study the temperature
fluctuations in homogeneous isotropic turbulence. This analysis leads
to the closure of the Corrsin equation and provides the statistics 
of temperature fluctuations.
The results, which represent a further application of the analysis presented in \cite{deDivitiis_1} and \cite{deDivitiis_2}, are here obtained in the case of self--similar velocity and temperature fluctuations, and can be so summarized:

\begin{enumerate}
\item
The energy equation, formally written using the referential coordinates and the Lyapunov analysis 
of the local deformation, allows to factorize the temperature fluctuation and to express it in the Lyapunov exponential form of the local deformation. 
\item
The finite--scale Lyapunov analysis provides an explanation of the physical mechanism of thermal energy  cascade and leads to the closure of the Corrsin equation. This is a non--diffusive closure equation
which expresses $G$ in terms of $f$ and $\partial f_\theta /\partial r$.
\item 
The closed Corrsin equation generates temperature spectra with different scaling laws, 
depending on $R$ and $Pr$.
In particular, for the proper values of $R$ and $Pr$, these spectra satisfy the Corrsin--Obukhov and  Batchelor scaling laws in opportune intervals of the wave--numbers.
\item
The Corrsin--Obukhov and Batchelor constants, here identified with the proposed theory, agree with the different source from the literature. 
\item
The PDF of $\vartheta_r$ and the corresponding dimensionless moments, are determined through a canonical decomposition of velocity and temperature in terms of proper random variables which describe the mechanism of energy cascade. This is a non--Gaussian PDF whose intermittency increases with $R$ and $Pr$, in agreement with experiments and simulations of prior studies.
\end{enumerate}

\bigskip

\bigskip

\section{\bf  Acknowledgments}

This work was partially supported by the Italian Ministry for the 
Universities and Scientific and Technological Research (MIUR).

\bigskip

\section{\bf Appendix}

For the sake of convenience, this section reports the main results of the finite scale
Lyapunov analysis obtained by \cite{deDivitiis_1} and \cite{deDivitiis_2}, which deal with the homogeneous isotropic turbulence.

\bigskip

\subsection{\bf Closure of the von K\'arm\'an--Howarth equation}

In fully developed isotropic homogeneous turbulence,
$f$ satisfies the von K\'arm\'an--Howarth equation (\cite{Karman38})
\bea
\ds \frac{\partial f}{\partial t} = 
\ds  \frac{K(r)}{u^2} +
\ds 2 \nu  \left(  \frac{\partial^2 f} {\partial r^2} +
\ds \frac{4}{r} \frac{\partial f}{\partial r}  \right) +\frac{10 \nu}{\lambda_T^2} f 
\label{vk-h}
\eea
the boundary conditions of which are
\bea
\begin{array}{l@{\hspace{+0.2cm}}l}
\ds f(0) = 1,  \\\\
\ds \lim_{r \rightarrow \infty} f (r) = 0
\end{array}
\label{bc0}
\eea
where  $\lambda_T \equiv \sqrt{-1/f''(0)}$ is the Taylor scale, and $u$ is the standard deviation of $u_r$, which satisfies the equation of the turbulent kinetic energy
\bea
\ds \frac{d u^2}{d t} = - \frac{10 \nu}{\lambda_T^2} u^2 
\eea
The function $K(r)$, related to the triple velocity correlation, 
represents the effect of the inertia forces and expresses the mechanism of energy cascade.
Thus, the von K\'arm\'an--Howarth equation provides the relationship between 
$\left\langle (\Delta u_r)^2 \right\rangle$ and $\left\langle (\Delta u_r)^3 \right\rangle$, where  $\Delta u_r$ is the longitudinal velocity difference.

The Lyapunov theory proposed in \cite{deDivitiis_1} leads to the
closure of the von K\'arm\'an--Howarth equation, and expresses  
$K(r)$ in terms of $f$ and $\partial f/\partial r$ 
\bea
\ds K(r) = u^3 \sqrt{\frac{1-f}{2}} \frac{\partial f}{\partial r}
\label{K}
\label{K closure}
\eea
$K(0)=0$ and this represents the property that $K$ does not modify the fluid kinetic energy (\cite{Karman38}, \cite{Batchelor53}).

\bigskip

\subsection{\bf Statistics of the longitudinal velocity difference}

Here, the results of \cite{deDivitiis_1}, dealing with the statistics of 
$\Delta u_r$ are recalled.
There, $\Delta u_r$ is represented in terms of centered random variables
\bea
\begin{array}{l@{\hspace{+0.2cm}}l}
\ds \frac {\Delta {u}_r}{\sqrt{\langle (\Delta {u}_r)^2} \rangle} =
\ds \frac{   {\xi_u} + \psi_u \left( \chi ( {\eta_u}^2-1 )  -  
\ds  ( {\zeta_u}^2-1 )  \right) }
{\sqrt{1+2  \psi_u^2 \left( 1+ \chi^2 \right)} } 
\end{array}
\label{fluc4}
\eea 
where $\psi_u$ is a function of $r$ and of the Taylor--scale Reynolds number
\bea
\psi_u({\bf r}, R) =  
\sqrt{\frac{R}{15 \sqrt{15}}} \
\hat{\psi}_u(r)
\label{Rl}
\eea
${\psi_{u 0}} = \psi_u(R,0)$, with  $\hat{\psi}_{u}(0) = 1.075$,
and $\chi \ne 1$ gives a nonzero skewness of $\Delta u_r$ (\cite{deDivitiis_1}).
Equation (\ref{fluc4}) arises from statistical considerations about the 
Fourier--transformed Navier--Stokes equations and  
expresses the internal structure of fully developed isotropic turbulence, 
where $\xi_u$, ${\eta_u}$ and $\zeta_u$ are independent centered random variables which exhibit 
gaussian PDFs $p(\xi_u)$, $p(\eta_u)$ and $p(\zeta_u)$ whose standard deviations are equal to the unity.

\bigskip

\subsection{\bf Self--Similarity in homogeneous isotropic turbulence}

Now, the results of the self--similarity  are briefly summarized.
These are based on the idea that far from the initial condition, 
the combined effects of energy cascade and viscosity act keeping $f$ and
$E(\kappa)$, similar in time for large values of wavelengths 
(\cite{Karman38, Karman49}).
This condition, applied to Eq. (\ref{vk-h}), 
leads to the following ordinary differential equation 
\bea
\begin{array}{l@{\hspace{-0.cm}}l}
\ds  \sqrt{\frac{1-f}{2}} \ \frac{d f} {d \hat{r}}  +
\ds \frac{2}{R}  \left(  \frac{d^2 f} {d \hat{r}^2} +
\ds \frac{4}{\hat{r}} \frac{d f}{d \hat{r}}  \right) + \frac{10}{R}  f = 0
\end{array}
\label{vk-h1}  
\eea
Into Eq. (\ref{vk-h1}), $f = f (\hat{r})$, where 
$
\ds  \hat{r} = r / \lambda_T(t)
$, therefore
$
{d^2 f}/{d \hat{r}^2}(0) = -1
$.
This similarity and the equation of $u^2$ lead to the
expressions of $u$ and $\lambda_T$
\bea
\begin{array}{l@{\hspace{-0.cm}}l}
\ds \lambda_T(t) = \lambda_T(0)\sqrt{1 +10 \nu \ t /\lambda_T^2(0) }, \ \ \ \ 
\ds u (t)= \frac{u(0)} {\sqrt{1+10 \nu \ t /\lambda_T^2(0) }}.
\end{array}
\label{ult}
\eea
As the solutions $f \in C^2 \left[0, \infty \right)$ with 
$d f/ d \hat{r}(0)=0$ tend to zero when $r \rightarrow \infty$, 
the boundary condition (\ref{bc0}) is replaced by the following conditions in the origin
\bea 
\ds f(0) = 1, \ \
\ds \frac{d f (0)} {d \hat{r}} = 0  
\label{bc3}
\eea
Therefore, the boundary problem represented by Eqs. (\ref{vk-h1}) and (\ref{bc0}), 
can be reduced to an initial condition problem written 
in the Cauchy's normal form
\bea
\begin{array}{l@{\hspace{+0.cm}}l}
\ds  \frac{d f}{d \hat{r}} =  F \\\\
\ds \frac{d F}{d\hat{r}} = -5 f -
\left( \frac{1}{2} \sqrt{\frac{1-f}{2}} R + \frac{4}{\hat{r}} \right) F
\end{array}
\label{vk-h2}  
\eea
the initial condition of which is 
\bea 
\ds f(0) = 1, \ \  F(0) = 0
\label{ic}
\eea

\bigskip

\end{document}